\documentclass[pre,a4paper,oneside,twocolumn,
nofootinbib]{revtex4}

\usepackage[utf8x]{inputenc}
\usepackage{comment}

\usepackage{amsmath}

\usepackage{hyperref}

\usepackage{float}
\usepackage{latexsym}
\usepackage{fancyhdr}
\usepackage{color} 
\usepackage{soul}
\usepackage{graphicx}  

\usepackage[dvipsnames]{xcolor}

\usepackage{rotating}

\usepackage[normalem]{ulem}

\usepackage{rotating, graphicx}
\usepackage{makecell, tabularx, multirow}

\usepackage{lscape}
\usepackage{longtable}

\begin{document}

\title{Mathematical models to explain the origin of urban scaling laws: a synthetic review}

\author{Fabiano L. Ribeiro\textsuperscript{*}}
\affiliation{Department of Physics (DFI), Federal University of Lavras (UFLA), Lavras MG, Brazil}
\author{Diego Rybski\textsuperscript{$\diamond$}	}
\affiliation{Potsdam Institute for Climate Impact Research -- PIK, Member of Leibniz Association, P.O.\ Box 601203, 14412 Potsdam, Germany}
\affiliation{Department of Environmental Science Policy and Management, University of California Berkeley, 130 Mulford Hall \#3114, Berkeley, CA 94720, USA}
\affiliation{Complexity Science Hub Vienna, Josefst\"adterstrasse 39, A-1090 Vienna, Austria}

\begin{abstract}
The quest for a theory of cities that could offer a quantitative and systematic approach to manage cities is at the top priority, given the challenges humanity faces due to the increasing urbanization and densification of cities.  If such a theory is feasible, then its formulation must be in a mathematical way. 
As a contribution to organizing the mathematical ideas that deal with such a systematic way of understanding urban phenomena, we present this material, concentrating on one important aspect of what recently has been called the new science of cities.
In this paper, we review the main mathematical models present in the literature that aim at explaining the origin and emergence of urban scaling.
We intend to present the models, identify similarities and connections between them, and find situations in which different models lead to the same output. 
In addition, we report situations in which some ideas initially introduced in a particular model can also be introduced in another model, generating more diversification and increasing the scope of the models.
The models treated in this paper explain urban scaling from different premises: from gravity ideas, passing through densification ideas and cites' geometry, to a hierarchical organization and socio-network properties.
We also investigate scenarios in which these different fundamental ideas could be interpreted as similar -- where the similarity is likely but not obvious. 
Furthermore, in what concerns the gravity idea, we propose a general framework that includes all gravity models analyzed as a particular case. 
\end{abstract}

\maketitle

\textbf{*} fribeiro@ufla.br

\textbf{$\diamond$} ca-dr@rybski.de



\section{Introduction}


For the first time in human history, the urbanised population surpasses the rural population, and United Nations estimate that until 2050 
more than 70\% of the people around the world will live in cities\footnote{\url{https://www.un.org/development/desa/en/news/population/2018-revision-of-world-urbanization-prospects.html}}.
To deal with all the problems that come with this urban intensification, like extreme density, traffic, infrastructure saturation, it is of urgency to develop a quantitative theory in order to understand the urban phenomena and to govern systematically our cities
\cite{Lobo2020,Bettencourt2010}. 
This theory will involve an interdisciplinary effort and could better predict scenarios to be explored by decision-makers and suggest new observations about the cities' growth and their organisation \cite{Batty2013}. 
This theory, if successful, can point out where the data is missing and what we need to measure to obtain a deeper understand of this phenomenon.
Besides, if we expect that this theory gives a quantitative description of cities, it must be formulated mathematically.
Then, trying to organise the mathematical ideas that treat such a systematic way to understand cities, we present this material, concentrating on one aspect that composes the \textit{new science of cities} \cite{Batty2013}: \textit{urban scaling}.

Scaling analysis proposes that some urban quantity, say  $Y$,  grows free-of-scale and non-linearly with the population size $N$ of a city, following the form 
$Y = Y_0 N^{\beta}$, where $Y_0$ is a constant and $\beta$ is the \textit{scaling exponent} \cite{bettencourt2007growth}. 
To a great extent, empirical evidence reveals three distinct scaling regimes. 

Variables related to socio-economic activities (e.g.\ GDP, Patents, AIDS cases) scale in a super-linear manner with the population size ($\beta> 1$).
Empirical evidence for economically and culturally different countries and also for different urban metrics suggest a numerical value of the scaling exponent around $\beta = 1.15$ for socio-economic variables \cite{bettencourt2007growth, joao_plosone2018}.  
It means the per-capita quantity of these socio-economic variables tends to increase with the size of a city -- the so-called increasing returns to
scale \cite{Sveikauskas1975}.
Typically, one large city generates more wealth than two cities of half the size together. We can intuitively say that \textit{the bigger the city is, the more wealth it generates} \cite{bettencourt2011bigger,Meirelles2020}.

On the other hand, variables associated with basic individual services (e.g.\ number of houses, water consumption) scale linearly with city population  ($\beta= 1$).
And infrastructure-related variables (e.g.\ electrical cables, number of gas stations) scale in a sub-linear manner ($\beta< 1$).
Empirical evidence suggests a numerical value of the scaling exponent around $\beta = 0.85$ for infrastructure variables \cite{bettencourt2007growth, Kuhnert2006, joao_plosone2018}. 
This means that bigger cities demand less infrastructure per-capita \cite{norman2006comparing}, which allows saying that: \textit{bigger cities do more with less} \cite{bettencourt2011bigger,Meirelles2020}.
There is also evidence for some kind of constraint on the numerical value of the scaling exponents, such that these super- and sub-linear exponents add up to $\approx 2$ \cite{bettencourt2013,ribeirocity2017}. 

But why does urban scaling emerge at all?
This paper tries to answer this question and focuses on works explaining non-linear urban scaling by some sort of model that goes beyond empirical characterization.
Many (but not all) of the models discussed here are built on the idea that socio-economic activity is the outcome of a multiplicative combination of population, geometry, and the probability of interaction between people.
The premise is that interaction, and consequently the exchange of knowledge, generates ideas that result in innovation, economic growth, increasing returns, and economies of scale.
In some models, the geometrical and network properties of the cities also play an essential role to explain the observed scaling laws, given that human interactions depend strongly on the city's spatial structure.
Natural Factors, such as rugged relief or the presence of physical barriers (mountains, rivers, lakes, etc.), promote or intensify the isolation of certain parts of the city.
In addition, artificial factors, e.g.\ the geometry of the street networks or the city's shape, should also affect human interactions.

We are aware that an enormous number of papers has been dedicated to present empirical and theoretical evidence about such urban scaling in the last years.
Of course, it will be not possible to organize in a single paper all the results and ideas contained in those works. 
We opt to present only models that explain or derive urban scaling properties as an emergent phenomenon, giving special attention to those that derive it mathematically. 

With the purpose of gaining insights from relating and comparing the models, we present them in a more straightforward manner than in the original publications.
The intention is to focus only on what is strictly essential to explain urban scaling quantitatively. Some models are rewriting using a different notation from the original to get homogenization and coherence among the models presented.
Most of the models' mathematical deductions are presented in a self-contained manner in this paper.
However, in some cases we opt to omit very extensive mathematical passages to preserve the text's dynamic and flux. 
In summary,  we intend to present a self-contained material that organizes different kinds of mathematical models.

\begin{figure*}
	\begin{center}
\includegraphics[width=0.80\textwidth]{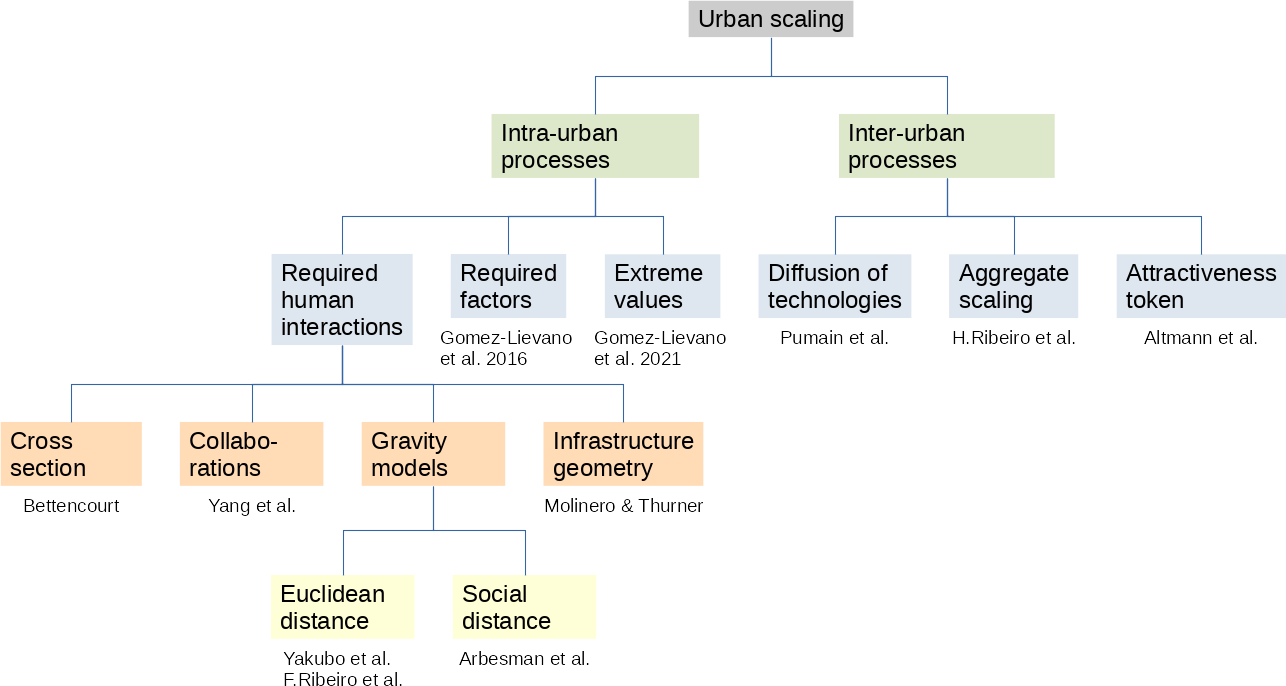}
\end{center}
	\caption{ \label{Fig_tax} 
	Taxonomy of the models explaining urban scaling. 
	Whether processes take place within or between cities is the first distinguishing factor.
	Many models are based on required human interactions within cities.
	Three gravitational models belong to this class.
}
\end{figure*}

The paper also aims at synergies from relating all those models. 
That is, what emerges from the interconnection between different models to explain urban scaling? Is it possible to see some common properties in different approaches? Also, writing the models in a standard notation allows us to identify what hypothesis and results they have in common.
Therefore, we organize the models in a taxonomy, identifying groups of models that share the same fundamental ideas, as organized in the Fig.~(\ref{Fig_tax}) and Tab.~(\ref{tab_models}). 
For instance, we find that a set of models differs only in how the interactions between people are estimated, i.e.\ how the probability of interaction is derived.  In the case of models based on gravity processes, 
this organization allows us to formulate a general framework with these models as particular cases.

The models that are presented here are divided into two categories: intra- and inter-city models. The intra-city models, which can be found in Sec.~(\ref{sec:intracmodels}), refer to models that consider only city internal factors to explain the scaling laws. In these models, the interaction between people, and how geometry and the city spatial distribution affects it, is the main component to explain the laws of scales. 
We also propose general formulations from which some models could be derived as particular cases. 
In this category, all gravity models represent special cases of a general formalization.

The second categories is about inter-city models, and it can be found in Sec.~(\ref{sec:intercmodels}). They consider the exchanging of some kind of information between cities to explain the scaling laws.  
Not all of the models are based on derivations as backbone and not all lead to emergence of urban scaling, but they have been included to better represent the somewhat less developed group of inter-city models.
The main mechanisms proposed to explain the scaling laws are:
Zipf's law, hierarchical organization, and interaction among people of different cities. 

\begin{figure}
	\begin{center}
		\includegraphics[width=\columnwidth]{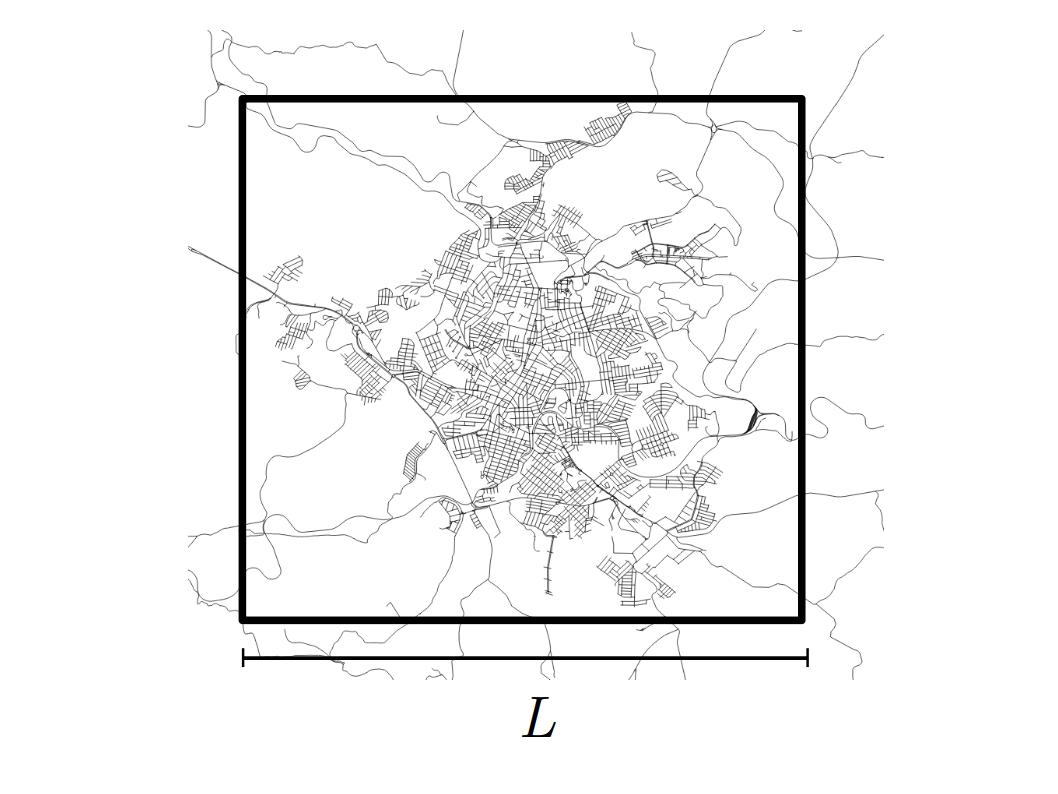}
		
	\end{center}
	\caption{ \label{Fig_area_circ} City of Lavras (Brazil) and its circumscribing area $A$, given by a square of size $L$,  that fully includes its build-up area $A_n$.
	}
\end{figure}

\section{Intra-city models}
\label{sec:intracmodels}

In the last years, we have seen a large number of works that consider (with empirical justifications) that urban variables are dependent on the population size $N$, without the necessity of incorporating other variables or even information from neighbouring cities.
Indeed, the main idea behind it is that the outcome of an urban variable is a  consequence of only city internal processes.
However, we know that cities are in constant interaction with each other, and the dependence of urban variables solely on the city size may be a manifestation of a successful first-order approximation. 
This section is dedicated to presenting the ideas of the mathematical models that explain urban scaling, and deriving an interpretation for the scaling exponent, using only endogenous factors.

We organize the intra-city models in two classes.
The first, covered in the following subsection, considers the interaction between people to explain non-linear urban scaling.  
The second class, addressed in Sec.~(\ref{sec:gomezlmodel}), considers 
that non-linear urban scaling emerges when a number of necessary complementary factors are available in the city.

\subsection{Required Human interaction}\label{sec_interaction}

The majority of publications on the topic consider human interaction as the primary mechanism to explain the origin of urban scaling. 
This fact obviously reflects in a larger space dedicated to this approach in the paper in hand.
We begin the description by introducing some quantities common to the models that belong to this category.
In addition, we present which properties a model based on interaction must have in order to be compatible with the empirical data.

\subsubsection{General framework of human interaction models} 
\label{sec:huminter}

Consider that the city is composed of $N$ individuals (population size of the city) that live in a \emph{circumscribing area} $A=L^2$, where $L$ is the size of the square that fully includes the \emph{build-up area} ($A_n$) of the considered city, cf.\ Fig.~(\ref{Fig_area_circ}).
When two individuals meet in the city they generate ideas that correspond to a quantity $g$ of socio-economic activity. 
For instance, $g$ could represent the amount of patents, the amount of wealth, etc., this encounter contributes to.
If each individual, say $i$, meets with $k_i$ persons, then the total wealth generated by these meetings is $g \cdot k_i$. 
The number $k_i$ can represent, depending on the model and without loss of generality, the \textit{number of contacts} (friends, colleagues, or random encounters) of this individual, or the number of interactions that he/she has in a specific period of time, or even the node degree in a complex network.

\begin{figure*}
\begin{center}
\includegraphics[scale=0.35]{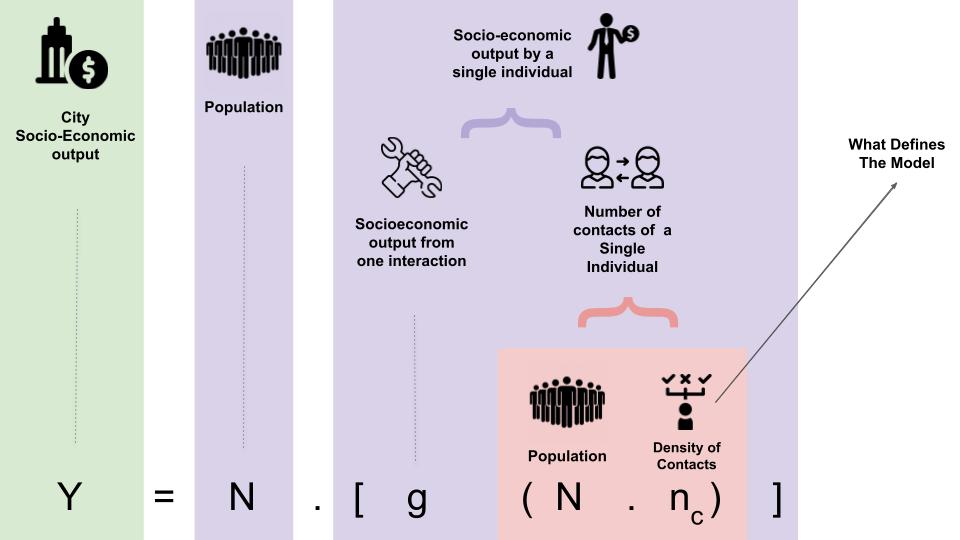}
\end{center}
\caption{ \label{Fig_diagrama}
Diagram illustrating the models that consider urban scaling as a result of knowledge exchange via human interactions.
The idea is that the total outcome $Y$ of a specific socio-economic variable can be understood as the sum of all individual socio-economic outputs of the city, that is $Y =  N y$. 
In turn, the individual socio-economic production $y$ is the result of the socio-economic output generated by a single interaction ($g$), multiplied by the total number of interaction of each individual ($k_i$). 
This idea yields the relation $Y = g N \langle k_i \rangle  =  g N^2 n_c$, where $n_c$, the average density of contacts, is what defines the models. 
}
\end{figure*}

What distinguishes the models presented below, is the way they propose to compute/determine the average number of contacts of the individuals, and  consequently the production of socio-economic wealth generated by these contacts. 
The considered models essentially obey the following description.
The city-wide total outcome $Y$ of a specific socio-economic variable can be  understood as the sum of all individual socioeconomic output. 
In turn, the \textit{individual socio-economic production}, namely $y$, is the result of the socio-economic output generated by a single interaction, multiplied by the total number of interaction of each individual, $y = g \cdot k_i$.
This idea is summarized in the diagram presented in Fig.~(\ref{Fig_diagrama}) and yields the relation  
\begin{equation}
Y = g N^2 n_c \, ,    
\end{equation}
where  
\begin{equation}\label{eq_nc}
n_c \equiv \langle k_i  \rangle/N   
\end{equation}
is the \textit{average density of contacts}. 
Essentially, the models differ in the way the authors propose to estimate this density.

If we consider that $g$ is scale independent (i.e.\ $g~\sim~N^0$, as suggested by \cite{bettencourt2013}), one obtains 
\begin{equation}\label{eq_Yp}
Y \sim N^2 n_c \,. 
\end{equation}
This means we are looking for $n_c$ that leads from Eq.~(\ref{eq_Yp}) to the empirical evidence
\begin{equation}\label{eq_Ysuper}
Y \sim N^{\hat{\beta}_{\text{super}}} \,,
\end{equation}
where $\hat{\beta}_{\text{super}} \equiv 1.15$ is (approximately) the \textit{empirical value} of the socio-economic scaling exponent. 
Equaling Eqs.~(\ref{eq_Yp}) and~(\ref{eq_Ysuper}), implies that the average density of contacts must follow the power-law 
\begin{equation}\label{eq_pint_res}
n_c \sim N^{\hat{\beta}_{\text{super}} - 2} \,.
\end{equation}

Let us also use $\hat{\beta}_{\text{sub}} \equiv 0.85$ to represent (approximately) the \textit{empirical value} of infrastructure scaling exponent. As it was suggested in \cite{bettencourt2013,ribeirocity2017}, there is some kind of complementarity between these two scaling exponents that can be expressed by the constraint 
\begin{equation}\label{Eq_sum2}
\hat{\beta}_{\text{super}} + \hat{\beta}_{\text{sub}} =  2\,. 
\end{equation}
It suggests that the exponent in Eq.~(\ref{eq_pint_res}) is, in fact, $\hat{\beta}_{\text{sub}}$, which allows us to write
\begin{equation}\label{eq_pint_res2}
n_c  \sim N^{-\hat{\beta}_{\text{sub}}} \,.
\end{equation}
This consideration is speculative at this points, but the following sections will provide some justification.
The expressions Eqs.~(\ref{eq_pint_res}) and~(\ref{eq_pint_res2}) represent a ``rule of thumb'' that the models based on interactions need to comply when quantitatively explaining scaling laws. 
The take home message here is that the density of contacts and consequently the probability of interaction of any proposed model based on interaction must 
result in Eqs.~(\ref{eq_pint_res}) and~(\ref{eq_pint_res2}) in order to be empirically consistent.

Table~\ref{tab_models} summarizes the models that will be presented in the next sections, organizing their main mechanism, as well as the scaling exponents predicted by them. 
Specific to the models based on interaction, 
all of them consist of computing the average number of contacts of the people. 
This quantity, in turn, will depend on the parameters used in each model.  
The following sections are dedicated to present these mathematical models in more details.

\begin{table*}
\begin{center}
\includegraphics[scale=0.66]{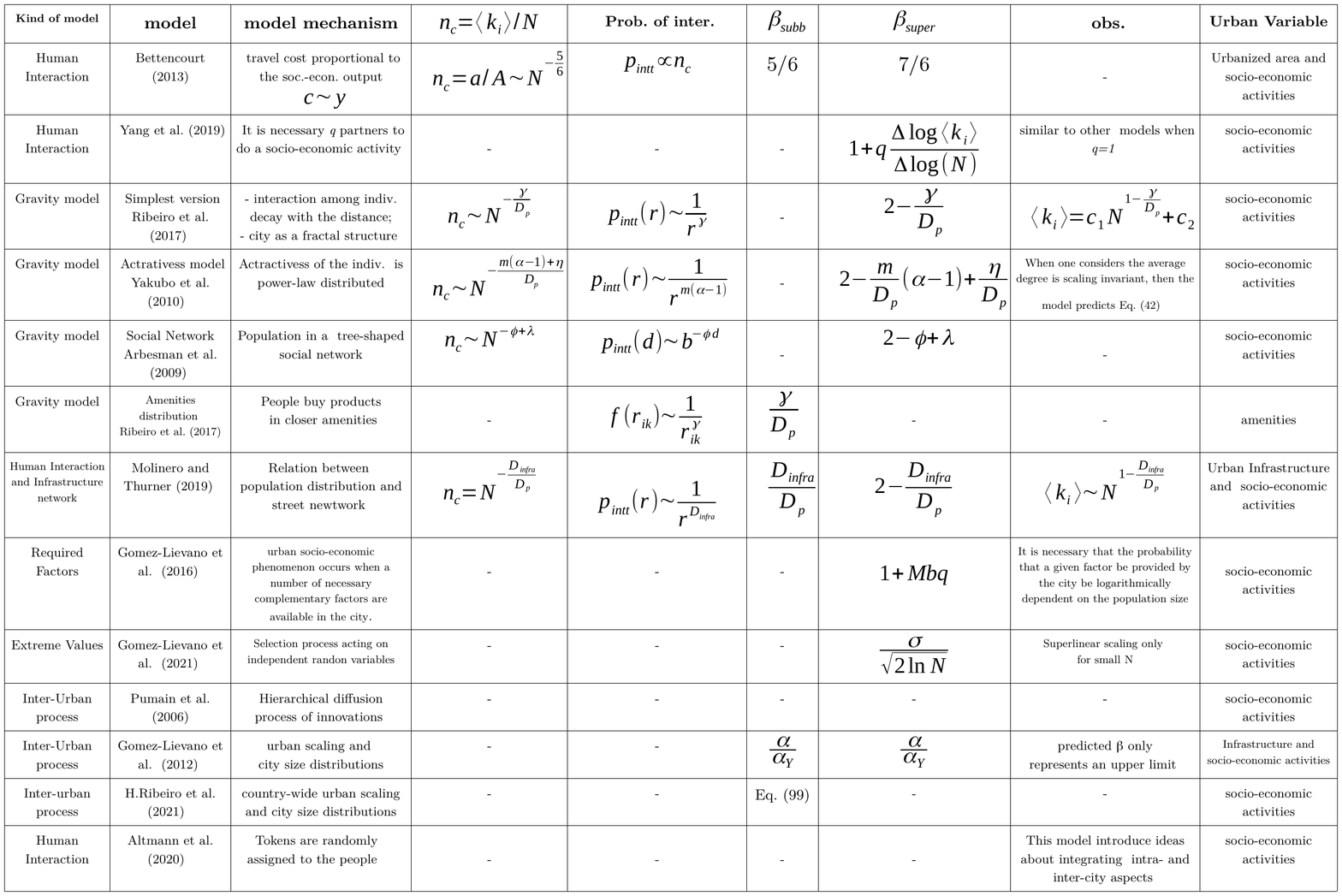}
\end{center}
\caption{\label{tab_models}
Overview of the main features of models that explain the non-linear urban scaling as a consequence of the interaction between people. Although each model considers different mechanisms, all of them compute the probably of interaction ($p_{\text{int}}$) between the people. This table also presents the predicted scaling exponents for each model.
}
\end{table*}

\subsubsection{Bettencourt model -- human interaction as cross section}
\label{sec_bettecourt}

Probably the most influential model  proposed to explain the origin of urban scaling is the one by Bettencourt \cite{bettencourt2013}.   
The model shares similarities with the concept of the \textit{cross section} as used in physics. 
It considers that each individual moves throughout the city prescribing, with an interaction radius $l_0$, a trajectory of length $l$.
It implies that this individual accesses in his/her trajectory an area $a = l_0 \cdot l$ of the city. 

The density of contacts can be considered to be the ratio 
\begin{equation}
n_c  = \frac{\text{Area accessible to the individual}}{\text{Area of the city}} = \frac{a}{A} 
\, . 
\end{equation}
For simplicity, at this point the city circumscribing area $A$ is considered.  
Using Eq.~(\ref{eq_Yp}) and keeping in mind that the area accessible is an intrinsic property of individuals, and therefore scale-independent ($a \sim N^0$), one obtains the relation 
\begin{equation}\label{eq_Ybetten}
Y \sim \frac{N^2}{A}
\, .
\end{equation} 

To determine $A$ as a function of $N$, Bettencourt assumes that each individual has a cost $c$ to move around in the city, and this cost is proportional to the length  $L$ of the city, that is $c \sim L$. 
Then the total transport cost of the city is $T = N c \sim N L$, and given that $A \sim L^2$, leads to
\begin{equation}\label{eq_T}
T \sim N \sqrt{A}
\, .
\end{equation}

Bettencourt also explores the hypothesis that the individual socio-economic production $y$ must be sufficient for each person to be able to travel through the city.
That is, $y$ must be sufficient to pay the transportation cost, which means $y \sim c$, ensuring a territorial unity of the city.
As a consequence of this hypothesis, and of $y=Y/N$, one has $T \sim Y$. 
This result, together with Eqs.~(\ref{eq_Ybetten}) and~(\ref{eq_T}), yields $A\sim N^{2/3}$, revealing a sub-linear regime between area and population of the city, as supported by empirical evidence 
\cite{Nordbeck1971}. 
However, the numeric value of the scaling exponent ($\approx 0.67$) is smaller than the one observed empirically ($\hat{\beta}_{\text{sub}} \approx 0.85$). 
Bettencourt attributes this discrepancy to the fact that the result is based on the circumscribing area ($A$) instead of the built-up area $A_n$.

To estimate $A_n$, he suggests to consider the average distance between individuals, say $\lambda$, and the density of individuals, say $\rho$.
These two quantities are related via $\rho = N/A = 1/\lambda^2$, which means that, on average, we have one single individual inside a square of size $\lambda$.
It implies that $\lambda = \sqrt{A/N}$, and as $A \sim N^\frac{2}{3}$, we obtain \begin{equation}\label{eq_free_path}
 \lambda \sim N^{-\frac{1}{6}}\, ,   
\end{equation}
and 
\begin{equation}
 \rho \sim N^{\frac{1}{3}} \, .    
\end{equation}
That is, the average distance between individuals decreases with the increase of city size. 
In other words, bigger cities are denser than smaller ones, as empirical studies suggest \cite{Nordbeck1971,Batty2011}.  

Finally one can say that $A_n \sim N \cdot \lambda$, and consequently, from Eq.~(\ref{eq_free_path}),  $A_n \sim N^{\frac{5}{6}}$, implying 
\begin{equation}\label{eq_An}
\beta_{\text{sub}} = \frac{5}{6} \, .
\end{equation} 
In relation to the socio-economic production, Eq.~(\ref{eq_Ybetten}) becomes $Y \sim N^2/A_n$, and consequently (inserting Eq.~(\ref{eq_An})) one gets $Y \sim N^{\frac{7}{6}}$, with
\begin{equation}\label{eq_Y_An}
\beta_{\text{super}} = \frac{7}{6} \, ,
\end{equation}
in accordance with empirical evidence \cite{bettencourt2007growth,joao_plosone2018}. 
For more components and details about this model and further development we refer to \cite{bettencourt2013,Bettencourtbook,Bettencourt2019a,Bettencourt2020a}.

In conclusion, the result Eqs.~(\ref{eq_An}) and~(\ref{eq_Y_An}) shows that Bettencourt's considerations  predict quantitatively the empirical scaling exponents. 
This is remarkable, given that the model is based on rather specific hypotheses. 
For instance, the model considers only the area as infra-structure variable
and does not take into account the number of amenities, 
that also scales sub-linearly with the population size \cite{Hidalgo2015a,Kuhnert2006,ribeirocity2017}. 
In addition, buildings are not taken into account by the model, which would expand the interaction range from a two-dimensional area to a three-dimensional volume.
The following models shed some light on this discussion.

\subsubsection{Yang et al.\ model -- required collaboration}

Yang et al. \cite{Yang2019} explain super-linear scaling as the result of the likelihood of finding required collaboration in the city, necessary for an undertaking.
Consider that the development of a certain prototype requires $q+1$ experts.
The case $q=0$ means that one single person can accomplish all the processes necessary for such activity. 
In the other mathematical models presented here, as the Bettencourt model and the gravity models presented in the next section, the activity and the productivity (that we call $g$) demands two persons, that is $q=1$.

The authors introduce $p_q \left( k_i \right)$ as the probability that an individual $i$ finds all $q$ collaborators required for an undertaking, among $k_i$ contacts.
The socio-economic production is then given by
\begin{equation}\label{eq_Y_yang}
Y \sim N p_q \left(  \langle k_i \rangle \right)
\, . 
\end{equation} 
where $\langle k_i \rangle$  is the average number of unique contacts for a person living in the city.
They show that $p_q \left( \langle k_i \rangle \right) \sim \langle k_i \rangle^q$ and consequently
\begin{equation}\label{Eq_Nn}
Y \sim N \langle k_i \rangle^q
\, .
\end{equation}
Let us first define
\begin{equation}
\Delta \log (N) \equiv \log N - \log N'  \, ,
\end{equation}
where $N$ and $N'$ are the populations of two different cities.
The same definition can be  applied to 
$\Delta \log (Y)$ and $\Delta \log \langle k_i \rangle$ in an analogous way. 
With this definition and given that $Y = Y_0 N^{\beta_{\text{super}}}$, where $Y_0$ is a constant (the intercept), one gets 
\begin{eqnarray}
\Delta \log (Y) &\equiv& \log Y - \log Y' \\
      &=&  \beta_{\text{super}}\log N + \log Y_0   -  \beta_{\text{super}}\log N' - \log Y_0 \, ,  \nonumber
\end{eqnarray}
and consequently 
\begin{equation}\label{eq_beta_yang}
    \beta_{\text{super}} =\frac{ \Delta \log (Y)}{ \Delta \log (N) } \, .
\end{equation}

Returning to Eq.~(\ref{Eq_Nn}),
applying the logarithm, and using the definition above yields
\begin{equation}
\Delta \log Y \sim  \Delta  \log N + q \Delta  \log \langle k_i \rangle
\, ,    
\end{equation}
in which dividing all the terms by $\Delta \log N $ and identifying Eq.~(\ref{eq_beta_yang}), one obtains the following expression for the scaling exponent
\begin{equation}\label{eq_beta_yang2}
\beta_{\text{super}} \sim 1 + q \frac{\Delta \log \langle k_i \rangle}{\Delta \log N} \, . 
\end{equation}

Some implications can be inferred from this result.
First of all, a necessary condition for super-linear scaling is that the
number of contacts is greater in larger cities, that is 
$\frac{\Delta \log \langle k_i \rangle}{\Delta \log N} \ge 0$, which is supported by empirical data \cite{Schlapfer2014}.
The second necessary condition for the super-linearity 
is that more than 1~person is necessary (that is $q > 0$) to implement the undertaking. 
In this sense, an undertaking that can be done alone ($q=0$) generates a liner scaling ($\beta_{\text{super}} \sim 1$) which falls into the category of individual-needs variables.
Eq.~(\ref{eq_beta_yang2}) implies also that if one does not need connections to realize something, the aggregate volume of this activity will scale linearly, without increasing returns to scale. 
In this sense, increasing returns can be attributed to enterprises that exceed $q=0$.
In contrast, Eq.~(\ref{eq_beta_yang2}) suggests that urban outputs requiring more participants should lead to more pronounced super-linear scaling.

\subsubsection{Gravity Models}\label{sec_gravity}

In this section a set of models is presented that explain urban scaling laws using the idea that the interaction between any pair of individuals within a city decays with the distance separating them -- similar to Newtonian gravity of two massive bodies. 
For a review about the application of gravity ideas in urban phenomena see \cite{Philbrick1973,Haynes1985,Barthelemy2019,BARTHELEMYbook}. 
Before presenting these models in more detail, let's define some quantities and introduce some concepts that are common to them.

\begin{figure}
	\begin{center}
\includegraphics[width=\columnwidth]{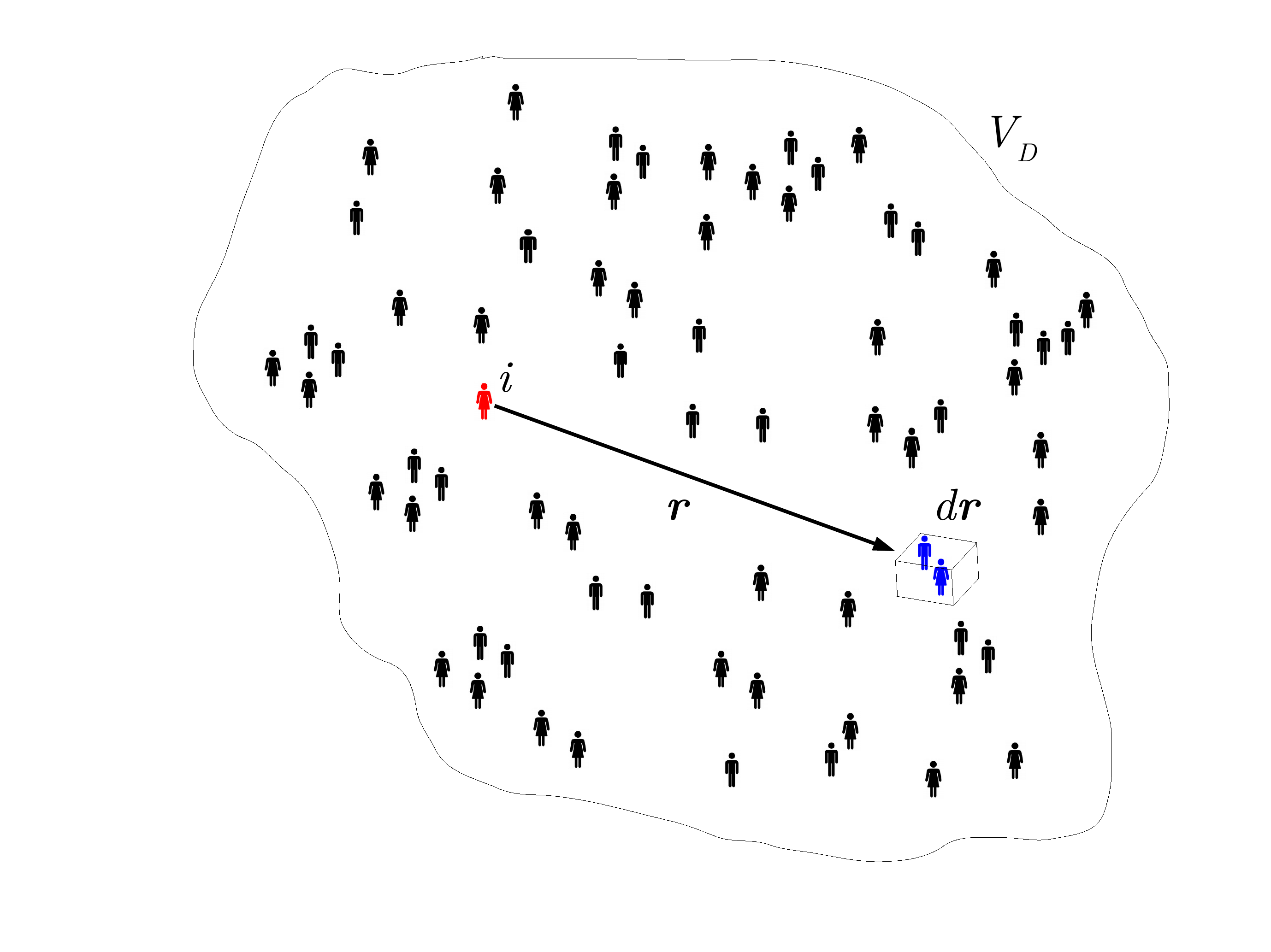}
	\end{center}
	\caption{ \label{Fig_populacao}
	Illustration of the particular case that the vector~$\mathbf{r}$ represents the Euclidean distance between the individual $i$ and the \textit{hyper-volume element} $d\mathbf{r}$ (as used in the context of gravity models). 
	All individuals inside this hyper-volume element are at distance $r = |\mathbf{r}|$ from $i$.
	The population (of the entire city) is completely embedded in the hyper-volume $V_D$.	
	 	 }
\end{figure}

First of all, we denote $dN(\mathbf{r})$ as the number of people inside a \textit{hyper-volume element} $d\mathbf{r}$ embedded in a $D$ dimensional space.
For instance, this hyper-volume element is an area if $D=2$; or a volume if $D=3$.  
Moreover, $\mathbf{r}$ is a vector directed from one particular individual, say $i$, to the individuals that are in the hyper-volume $d\mathbf{r}$. 
This vector can be interpreted in two ways, (i) as \textit{position vector} in $D$ dimensions, and consequently its modulus is the Euclidean distance $r$, conform illustrated in Fig.~(\ref{Fig_populacao}); or (ii) as non-Euclidean distance that separates any two individuals inside a complex network.

With the definition of theses quantities, one can compute, for instance, the total number of individuals in the city as $N = \int dN(\mathbf{r}) d\mathbf{r}$ or the \textit{density of individuals} at $\mathbf{r}$ as $\rho(\mathbf{r}) = dN(\mathbf{r}) / d\mathbf{r}$, which characterize how the population arranges itself in space. 
An empirical study considering such heterogeneous spatial distribution of people is presented in \cite{Dong2020}.

The gravity idea enters when we considering the probability, say $p_{\text{int}} (\mathbf{r})$, of the $i$-th individual interacting with -- or to be a contact of -- someone who is at $\mathbf{r}$. 
Then $p_{\text{int}} (\mathbf{r}) dN(\mathbf{r})$ is the (average) number of contacts that this individual has in $\mathbf{r}$, and consequently the total number of contacts of this individual, say $k_i$, will be given by the integral 
\begin{equation}\label{eq_Nci_grav}
k_i = \int p_{\text{int}} (\mathbf{r}) 
dN(\mathbf{r}) = \int p_{\text{int}} (\mathbf{r}) \rho (\mathbf{r}) d\mathbf{r} \, .  
\end{equation}

Moreover, let's denote $g(\mathbf{r})$ as the socio-economic production generated by the interaction between $i$ and another individual located at $\mathbf{r}$.
It is plausible to assume that the interaction between two more distant individuals can be more productive than the interaction between closer individuals  (``The strength of weak ties'' 
\cite{Granovetter1973,Arbesman2009}), since distant individuals are exposed to different experiences. 

With these considerations, it makes sense to interpret $p_{\text{int}} (\mathbf{r}) dN(\mathbf{r}) \cdot g(\mathbf{r})$ as the socio-economic production generated by the interaction between $i$ and all the other individuals at $\mathbf{r}$. 
Then the total socio-economic production of this individual is
\begin{equation}
y_i =  \int p_{\text{int}} (\mathbf{r}) \rho (\mathbf{r})  g(\mathbf{r}) d\mathbf{r}  \, . 
\end{equation}
Finally, the total socio-economic production of the city, that is $Y = N y_i$, can be written as  
\begin{equation}\label{eq_geral_grav}
Y =  N \int p_{\text{int}} (\mathbf{r}) \rho (\mathbf{r})  g(\mathbf{r}) d\mathbf{r} \, .  
\end{equation}
This is the generic formulation of the models based on the gravity idea. 
What differentiates these models is the way they propose to define:
\begin{enumerate}
    \item the metric associated with the vector  $\mathbf{r}$;
    
    \item the probability of interaction $p_{\text{int}} (\mathbf{r})$ and its dependence (decay) with $\mathbf{r}$; 
    \item the spatial distribution of the population, characterized by $\rho (\mathbf{r})$;
    \item and the socio-economic production $g(\mathbf{r})$ generated per encounter/contact.
\end{enumerate}

In the following we present the gravity models, from the simplest version to more complex ones.
Table~\ref{tab_models} summarizes the main findings of these models.

\subsubsection*{F.Ribeiro et al.\ model -- simple gravity}
\label{sec_gravity_simplest}

The model studied by Ribeiro et al.\ \cite{ribeirocity2017} considers that the probability of interaction between two individuals decays with the \emph{Euclidean distance} $r$ that separates them according to a power-law
\begin{equation}\label{eq_f_r}
p_{\text{int}}(r) = \frac{1}{r^{\gamma}}
\, , 
\end{equation}
where $\gamma$, the \emph{decay exponent}, is a parameter of the model which measures the \textit{interaction range}.
Some empirical evidence that support the hypothesis Eq.~(\ref{eq_f_r}) can be found in \cite{Goldenberg2009,Herrera-Yague2015,Scellato2010}.

The model also considers that the \textit{population space distribution} forms a fractal structure with fractal dimension $D_p$ embedded in a hyper-volume with Euclidean dimension $D$ (holding $D_p \le D$).
This assumption allows to write the density of individuals at $\mathbf{r}$  as 
\begin{equation}\label{eq_rho}
\rho(\mathbf{r}) = \frac{\textrm{number of individuals}}{\textrm{hyper-volume}} = \rho_0 \frac{r^{D_p}}{r^D} = \rho_0 r^{D_p - D}
\, ,
\end{equation} 
where $\rho_0$ is a constant.
Finally, they consider that all interactions have the same socio-economic production, that is
\begin{equation}\label{eq_ribeiro_g}
g(\mathbf{r}) = \text{const} \, . 
\end{equation}

Combining all assumptions [Eqs.~(\ref{eq_f_r}), (\ref{eq_rho}) and~(\ref{eq_ribeiro_g})], the total socio-economic production of the city can be determined by solving the integral Eq.~(\ref{eq_geral_grav}). 
This can be done by transforming the integration element from Cartesian to hyperspherical coordinates, that is using
\begin{equation}\label{eq_dr}
 d\mathbf{r} = r^{D-1}dr d\Omega \, ,   
\end{equation}
where $d\Omega$ is the \emph{solid-angle}, which leads to   
\begin{equation}\label{eq_I_c1c2}
Y = c_1 N^{ 2 - \frac{\gamma}{D_p}} + c_2 N
\, ,
\end{equation}
where $c_1$ and $c_2$ are constants. 
A similar calculus using polar coordinate is presented in \cite{Li2017}. Using similar ideas one can calculate the average number of contacts using Eq.~(\ref{eq_Nci_grav}), that yields to $\langle k_i \rangle = c_1 N^{1- \frac{\gamma}{D_p}}+c_2$,  and the density of contacts
using Eq.~(\ref{eq_nc}), which yields $n_c \sim N^{-\frac{\gamma}{D_p}}$. 

This result allows the following interpretation. 
The case $\gamma > D_p$ characterizes a \emph{short-range interaction regime} where the linear term in Eq.~(\ref{eq_I_c1c2}) dominates for sufficiently large $N$.
That is, the system converges to $Y \sim N$, i.e.\ a linear relation between urban metric and population. 
The case $\gamma > D_p$ characterizes a \emph{long-range interaction regime} where the non-linear term in Eq.~(\ref{eq_I_c1c2}) dominates for sufficient large $N$. 
That is, the system behaves in a super-linear way, characterized by $Y \sim N^{2 - \frac{\gamma}{D_p}}$.
This means when there are long-range interactions between the individuals, then the super-linear scaling exponent will be given by 
\begin{equation}\label{eq_beta}
\beta_{\text{super}} = 2 - \frac{\gamma}{D_p}
\, .
\end{equation}
This result suggests also that the scaling exponent is determined by the ratio of two geometrical parameters, the decay exponent and the population fractal dimension.
Some recent works \cite{Perez-Garcia2020,ribeiro_tumor2017} applied the same approach adopted here to
study tumour growth, reaching similar results. 
It allows some kind of analogy between urban systems and the interaction of cancer cells.

According to this model, the super-linearity of the socio-economic variables, expressed by Eq.~(\ref{eq_beta}), is a consequence of the integrity of the city, in the sense that the super-linear behavior of the socio-economic activity should only appear when there are interactions within the entire city (long-range regime). Otherwise, if the city is formed by isolated regions (short-range regime), the number of interaction and consequently the socio-economic metrics will depend linearly on the population size, without increasing returns to scale.
It is interesting to note that this result is in accordance with an argument used by Bettencourt
(see Sec.~\ref{sec_bettecourt}), namely that the per-capita socio-economic production must be sufficient to pay the transportation cost ($y \sim c$), ensuring a territorial unity of the city. 
Remarkably, two different approaches use the same argument to explain urban scaling, i.e.\ interconnection between the parts that constitute the city.

In addition, the result Eq.~(\ref{eq_beta}) also allows us to conclude that the smaller the $\gamma$, that is, the larger the access of the people to more distant parts of the city, the more pronounced is the socio-economic scaling.
That is, the larger the region of people's access, e.g.\ due to an efficient transport system, the better the city's socio-economic metrics and the more pronounced are the increasing return to scale\footnote{It is worth mentioning that a larger $\beta_{\rm{super}}$ does not automatically imply a wealthier urban system. It can also be a result of an economic imbalance between smaller and larger cities.}.
Obviously, an efficient transport system only represents a necessary condition to increasing return to scale. 
Further conditions -  as the presence of influential people integrating distant parts of the city or the interaction between socially distant people -  are discussed in the following sections (other gravity models).
The decay exponent $\gamma$ rather represents a compound value of the influence of distance \cite{Cliff1974,Couclelis1996}. 
This leads to the question about the role of information and communication technology (ICT).
With the establishment of the Internet and e.g.\ video conference systems, physical vis-\`a-vis meetings might become less important, which would reduce the influence of the distance
\cite{Couclelis1996, Lengyel2015}. 
However, a temporal evolution of $\gamma$ remains to be proven empirically.

With the following models, and with the introduction of further concepts, we present some insights and possible interpretation for this parameter $\gamma$.

\subsubsection*{Yakubo et al.\ model of individual attractiveness}\label{sec_yakubo}

Yakubo et al.\ \cite{Yakubo2011,yakubo2014} consider an additional ingredient, namely that people exhibit different attractiveness to one another depending on how influential an individual is.
To model this aspect, the authors consider a set of random variables $\{x_i\}_{i=1..N}$, each one associate to a give individual and following a power-law distribution (a pdf) 

\begin{equation}\label{eq_yako_Sx}
s(x) \sim x^{-\alpha} \, ,   
\end{equation}
where $\alpha$ is a parameter of the model. 
The higher the value of $x$ the more attractive is the individual. 

Any two individuals, say $i$ and $j$, are connected to each other 
if 
\begin{equation}\label{eq_yakubo_xixj}
\frac{x_i x_j}{r_{ij}^{m}} > \Theta
\, , 
\end{equation}
where $\Theta$ is a threshold constant, $r_{ij}$ is the Euclidean distance between them, and the exponent $m$ is a parameter of the model.
The probability that the $i$-th individual is connected to another individual at distance $r$ can be computed by 
\begin{equation}\label{eq_pint2_yak}
p_{\rm{int}} (r) =   \left\{ \begin{array}{ll}
1, & \textrm{if $r \le \xi$} \\
 &  \\
\int_{x> \Theta r^{m}/x_i} s(x) dx,  &  \textrm{ if $r > \xi$}
\end{array} \right.
\, ,
\end{equation}
where $x = \Theta r^{m}/x_i$ is the lower limit of the $x$ values necessary for an individual at $r$ to interact with $i$, and
\begin{equation}\label{eq_xi_theta}
\xi \equiv \left( \frac{x_{\rm{min}}^2}{\Theta} \right)^{\frac{1}{m}}
\end{equation}
is a distance below which any two individuals are connected regardless of $x$. 
Here $x_{\rm{min}}$ is the smallest value assumed by $x$, i.e.\ $x_{\rm{min}} \equiv {\rm min}\{ x_i \}$.
A condition for convergence of the integral in Eq.~(\ref{eq_pint2_yak}) is $\alpha>1$, 
and if this is the case, then the solution of this integral, using the distribution~(\ref{eq_yako_Sx}),  is 
\begin{equation}\label{eq_p_yakubo}
p_{\text{int}}(r) \sim  \frac{1}{r^{m (\alpha -1)}}
\, , 
\end{equation}
for $r> \xi$. 
Note that if we identify 
\begin{equation}\label{eq_gamma_alpha}
 \gamma = m (\alpha -1)   \, ,
\end{equation}
then we recover Eq.~(\ref{eq_f_r}), as used in the Ribeiro et al.\ model (Sec.~\ref{sec_gravity_simplest}). 

The authors also consider a more generic shape for the socio-economic production generated by each interaction, namely
\begin{equation}\label{eq_p_yakubo_g}
g(r) \sim r^{\eta} \, ,
\end{equation}
where $\eta$ is a parameter, in the sense that the productivity increases with the distance when $\eta> 0$ and decreases with the distance when $\eta < 0$; $\eta = 0$ means that all connections have the same socio-economic contribution as considered in the previous model.
The authors also consider that the \textit{population spatial distribution} is a fractal structure with dimension $D_p$ and therefore Eq.~(\ref{eq_rho}) is also valid in this context. 

Combining all ingredients of the model, i.e.\ using Eqs.~(\ref{eq_p_yakubo}), (\ref{eq_p_yakubo_g}), and~(\ref{eq_rho}), it is possible to compute the average degree (average number of contacts), namely $\langle k_i \rangle$, via 
\begin{equation}
\langle k_i \rangle = \int \int \rho(r) p_{\rm{int}}(r) s(x) dx dr \, , 
\end{equation}
and the socio-economic output with Eq.~(\ref{eq_geral_grav}). 
It results, respectively, in 
\begin{equation}\label{eq_ki_yak}
   \langle k_i \rangle =  c_1 \xi^{D_P} + c_2 \xi^{m(\alpha -1)} N^{ 1- \frac{m(\alpha-1)}{D_P}}   
\end{equation}
and 
\begin{equation}\label{eq_Y_yak}
Y = c_3 \xi^{\eta +D_P} N +  c_4 \xi^{m(\alpha-1) } N^{2 + \frac{\eta - m(\alpha-1)}{D_P}} \, ,
\end{equation}
where $c_1, c_2, c_3$ and $c_4$ are constants. 

For the interpretation of this result, it is necessary to distinguish two situations concerning the distance $\xi$ (or $\Theta$ parameter in Eq.~(\ref{eq_xi_theta})) and its scaling properties.
If $\xi$ is scale-invariant (i.e.\ $\xi \sim N^0$), and $N$ sufficiently large, then
\begin{enumerate}
\item $Y \sim N$ when $m > (D_p + \eta)/(\alpha -1)$, i.e.\ the first right-hand term in Eq.~(\ref{eq_Y_yak}) dominates, which characterize a short-range interaction regime (see Sec.~\ref{sec_gravity_simplest});  and
\item  $Y \sim N^{2 + \frac{\eta - m(\alpha-1)}{D_p}}$ when $m < (D_p + \eta)/(\alpha -1)$, i.e.\ the second right-hand term in Eq.~(\ref{eq_Y_yak}) dominates, characterizing a long-range interaction. 
\end{enumerate}
This means, the super-linear scaling behaviour happens in a long-range kind regime, with exponent
\begin{equation}\label{eq_beta_Yak1}
\beta_{\text{super}} =2- \frac{m}{D_p} (\alpha -1) + \frac{\eta}{D_p}
\, .
\end{equation}
According to the model, the super-linearity of the socio-economic scaling exponent can occur even when the productivity generated by the interaction is independent of the distance (i.e.\ when $\eta = 0$). 
In fact, the main factor that controls this super-linearity is the ratio between the interaction range (expressed by $\gamma = m(\alpha-1)$) and the fractal dimension of the city ($D_p$), as it was already suggested in the previous section.
Indeed the Yakubo et al.\ model and the gravity model studied by Ribeiro et al.\ are equivalent when $\eta = 0$ and $\alpha = 2$. 

In addition, the result Eq.~(\ref{eq_gamma_alpha}) gives a more fundamental explanation for the parameter $\gamma$.
It suggests that this parameter, which controls the interaction range, depends not only on the geometric properties -- expressed by the parameter $m$ 
-- but also on the degree of influence of the people who compose the city, expressed by the parameter $\alpha$.
Moreover, when the parameter $\alpha$ is sufficiently large, which represents the situation where the influence is distributed around a typical value, then a short-range interaction regime is observed. 
It corresponds to a more homogeneous population in terms of influence. 
A larger $\alpha$ -- which can also be thought of as an absence of a concentration of influence -- leads to a smaller $\beta_{\text{super}}$, i.e.\ it reduces the increasing returns to scale.  
Conversely, if $\alpha$ is sufficiently small, representing the situation where some people exert a considerable influence on the population, then a long-range interaction regime is observed, where the city behaves in a more integrated way.
To sum up, the result Eq.~(\ref{eq_beta_Yak1}) suggests that to improve the urban socio-economic metrics (larger $\beta_{\text{super}}$), it is essential not only to provide good access to other parts of the city -- as it was discussed in the previous subsections -- but also to have influencers in the population who can establish interaction between distant parts of a city. 

\begin{figure}
	\begin{center}
	\includegraphics[scale=0.4]{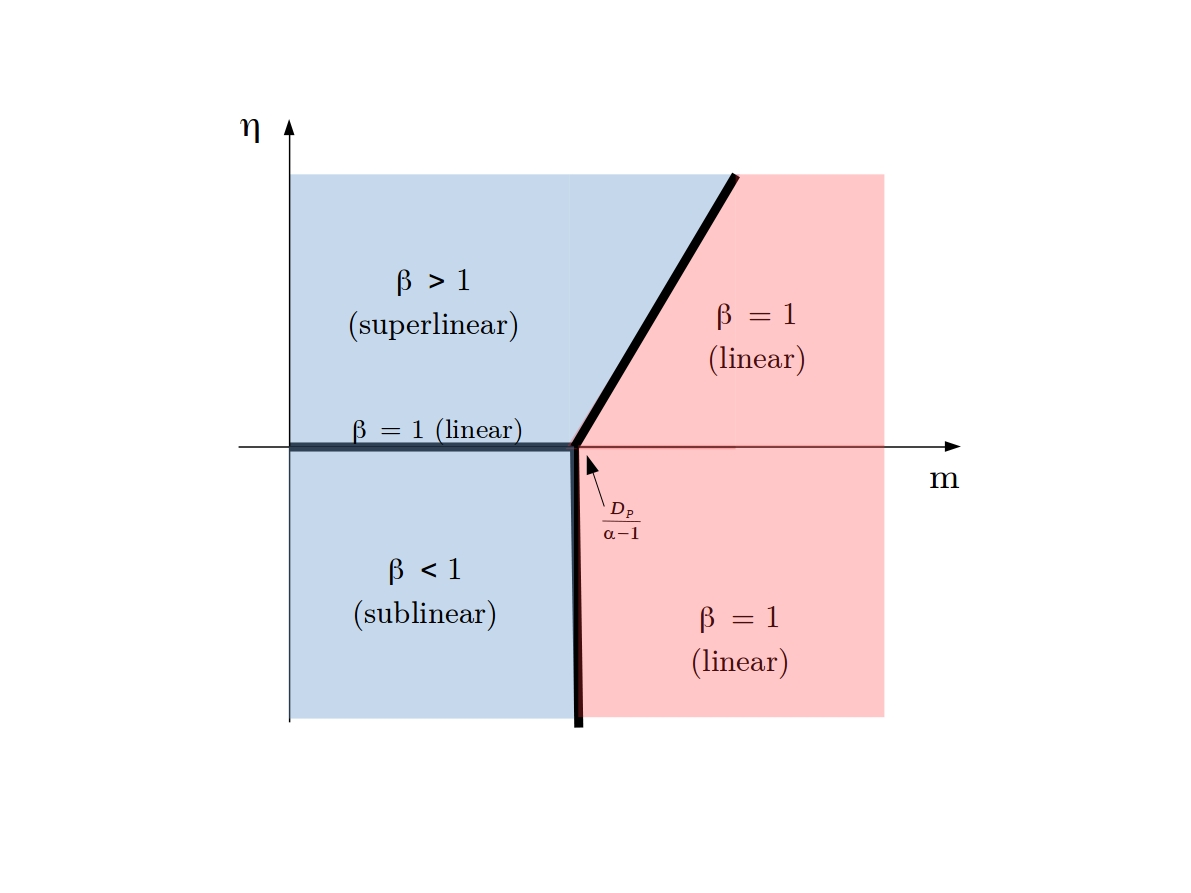} 
\end{center}
	\caption{ \label{Fig_Yakubo} 
	Phase diagram of the possible regimes (super-linear, sub-linear, and linear) according to the Yakubo et al.\ model of individual attractiveness, Eq.~(\ref{eq_pint_yak}), and the model parameters $\eta$, $m$ , $D_p$ and $\alpha$.  The non-linearity (sub or super-linear) only happens when $\eta \ne 0$.  The diagram also presents the parameter configuration that yields a long-range interaction regime (blue filling) and a short-range interaction regime (light-red filling). 	The super-linearity only occurs in the presence of long-range interaction and with $\eta$ positive;  that is, when the productivity among pairs increases with the distance.
	Source: modified after Yakubo et al. \cite{yakubo2014}.}
\end{figure}

However, the result of the scaling exponent changes drastically when we consider an average degree that is scale-invariant, i.e.\ $\langle k_i \rangle \sim N^0$, as proposed originally \cite{yakubo2014}.
According to Eq.~(\ref{eq_ki_yak}), this implies that $\xi$ scales with the population size $\xi \sim \langle k_i \rangle^{\frac{1}{D_p}}$ when $m> D_p/(\alpha -1)$ and $\xi \sim N^{\frac{1}{D_p} - \frac{1}{m(\alpha-1)}}$ otherwise.
Inserting such result in Eq.~(\ref{eq_Y_yak})  yields the following scaling exponents.
\begin{equation}\label{eq_pint_yak}
\beta  =   \left\{ \begin{array}{ll}
1, & \textrm{if  $m(\alpha -1) \le D_p$} \\
& \textrm{and $m(\alpha -1) \le D_p + \eta$} \\
& \\
2- \frac{m (\alpha -1) - \eta }{D_p}, 
& \textrm{if  $D_p \le  m(\alpha -1) \le D_P+\eta$} \\
& \\
2 + \frac{\eta}{D_p} - \frac{D_p + \eta}{m(\alpha -1) }, & \textrm{if  $m(\alpha -1)< D_p$} \\
& \textrm{and $m(\alpha -1) > D_p + \eta$}
\\
& \\
1+ \frac{\eta}{D_p},  & \textrm{if  $m(\alpha -1)< D_p$} \\
& \textrm{and $m(\alpha -1) < D_p + \eta$}
\end{array} \right.
\end{equation}
Figure~(\ref{Fig_Yakubo}) synthesizes these results, revealing many possibilities of regimes (sub-linear, super-linear, and linear) for the scaling exponents according to the parameters of the model.
It is important to note the role of the parameter $\eta$ in this context.
The non-linearity (super- or sub-linear regimes) only happens for $\eta \ne 0$. 
The value of this parameter can also change the regimes, from sub-linear ($\eta <1$) to super-linear ($\eta>1$). 
The parameter $\eta$ also changes the region of parameters that we are interpreting as a long-range interaction (blue part of Fig.~\ref{Fig_Yakubo}) and the short-range interaction (light-red part of Fig.~\ref{Fig_Yakubo}). 
The main results of this model are summarized in the Tab.~(\ref{tab_models}).

\subsubsection*{Arbesman et al.\ model -- Tree-shaped social network}

Up to now, the models that we presented are based on the idea that the interaction between the people needs to overcome geographical distance.
However, the work developed by Abersman~et~al.~\cite{Arbesman2009} shows that urban super-linear scaling can also emerge when a hierarchically organized social network is considered. 

\begin{figure}
	\begin{center}
		\includegraphics[scale=0.27]{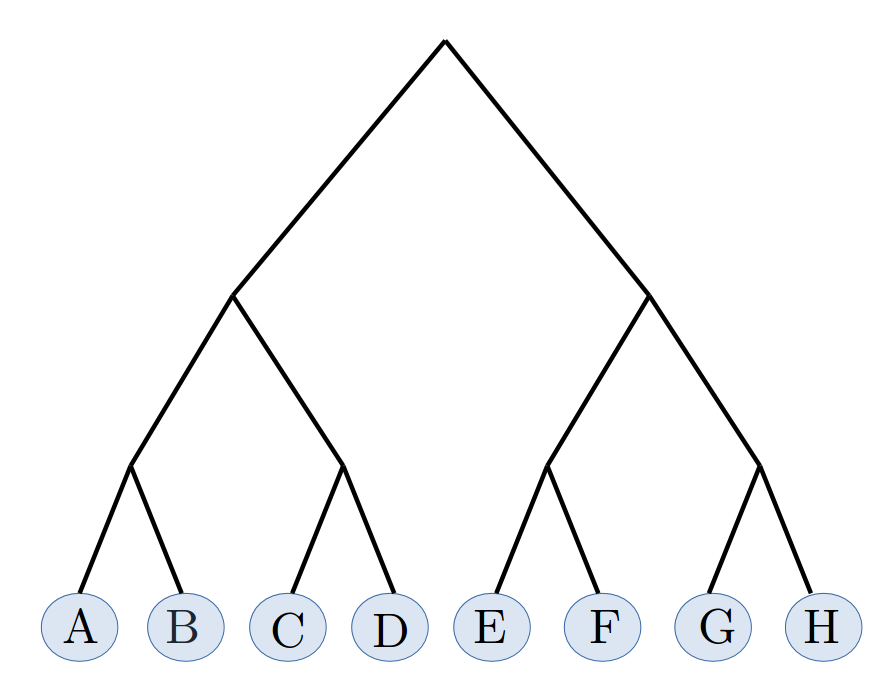}  
\end{center}
	\caption{ \label{Fig_tree}  
	Tree-shaped social-network -- as used by the Arbesman et al.\ model -- composed by $N=8$ individuals, named $A, B, ...,H$,  disposed in $N$ leaves. In this particular case, all branches splits in $b=2$ other branches.
	The distance $d$ between any two individuals in this network is the height of their lowest common ancestor. For instance, the social distance between $A$ and $B$  ($AB$) is $d=1$; $AC$ and $AD$ are $d=2$; $AE$, $AF$, $AG$ and $AH$ are $d=3$. Source: \cite{Arbesman2009}.}
\end{figure}

The authors propose that the population is organized in a tree-shaped social network, as the one sketched in Fig.~(\ref{Fig_tree}).
The $N$ individuals of the population are in the $N$ leaves of this tree, and  every branch in this hierarchical network splits into $b$ new branches. 
The social distance $d$ between two individuals is defined as the height of their lowest common ancestor.
In relation to the general gravitational framework that was presented in the beginning of this section, the distance vector becomes a scalar, that is $\mathbf{r} \to d$, and it does not represent the physical distance, but rather the social distance.

One can demonstrate that the \textit{number of leaves} that are at social distance $d$ from a given individual, say $N(d)$, grows exponentially
\begin{equation}\label{eq_net_Nd}
N(d) = b^d \, .
\end{equation}
The exponential structure of this relation motivates to model the other quantities necessary to compute Eq.~(\ref{eq_geral_grav}) with other exponential functions. 
For instance, the authors considered that the probability of interaction between two nodes drops off exponentially with the network distance as
\begin{equation}\label{eq_net_pint}
p_{\text{int}}(d) \sim  b^{-\phi d}
\, ,
\end{equation}
where $\phi$ is the parameter that controls the range of interaction inside this network. 
Finally, the authors propose that the social productivity of the interaction between two individuals also depends exponentially on the social distance between them
\begin{equation}\label{eq_net_g}
g(d) = b^{\lambda d} \, .
\end{equation}
Here $\lambda$ is a parameter, in the sense that the productivity increases with the social distance when $\lambda>0$ and decrease with the social distance when $\lambda<0$.
All interactions have the same productivity when $\lambda=0$.
This parameter is similar to the parameter $\eta$ in the context of the Yakubo et al.\ model of individual attractiveness (see Sec.~\ref{sec_yakubo}). 

We determine the total productivity inserting these tree relations Eqs.~(\ref{eq_net_Nd}), (\ref{eq_net_pint}), and~(\ref{eq_net_g}) in the general relation Eq.~(\ref{eq_geral_grav}). 
Given that $d$ is a discrete variable, implying $d \mathbf{r} \to \Delta d =1$, the integral in Eq.~(\ref{eq_geral_grav}) becomes the sum
\begin{equation}
Y = N \sum_{d=1}^{\log_b N } b^{-\phi d} b^d b^{\lambda d} \, ,
\label{eq:arbessum}
\end{equation}
where $\log_b N$ is the maximum social distance in the network.  
The sum in Eq.~(\ref{eq:arbessum}) is in fact a geometric progression which can be solved analytically, yielding $Y \sim N^{2 - \phi + \lambda}$, from which 
\begin{equation}\label{eq_net_beta}
\beta_{\text{super}} = 2 - \phi + \lambda
\end{equation}
follows.

This result shows that the socio-economic scaling exponent is larger when socially distant people interact
(characterized by smaller values of $\phi$ and positive values of $\lambda$).
This means, the city improves its socio-economy when the interaction among socially different people is possible -- in a similar way as it was discussed previously in the context of geographic distance. 
As argued in \cite{Arbesman2009},
``rich interconnectivity between communities creates better cities'', implying that socially distant ties can be a socio-economic force.

For the sake of completeness, given the probability of interaction Eq.~(\ref{eq_net_pint}), together with Eq.~(\ref{eq_net_Nd}), one can calculate the average number of contacts of this model using Eq.~(\ref{eq_Nci_grav}), which yields $\langle k_i \rangle = N^{1- \phi +\lambda}$, and the density of contacts using Eq.~(\ref{eq_nc}), which yields $n_c \sim N^{-\phi +\lambda}$.
These results are summarised in the Tab.~(\ref{tab_models}).

\subsubsection*{Connection between Euclidean and social distance}

We have discussed the gravity ideas in two versions, considering physical distance and social distance. 
However, we would also like to proof that under certain circumstances these two versions can be understood as equivalent.

First of all, if we consider that the two models 
are compatible, then the probability of interaction between any two individuals must be the same, that is
\begin{equation}\label{eq_pint_conec}
p_{\text{int}} \sim  b^{-\phi d } \sim  r^{-\gamma} \,
\end{equation}
if we combine Eqs.~(\ref{eq_f_r}) and~(\ref{eq_net_pint}).
Moreover, if both model-versions are compatible, then also the scaling exponent must be the same, i.e.\ combining Eqs.~(\ref{eq_beta}) and~(\ref{eq_net_beta}), and considering $\lambda =0$ without loss of generality leads to 
$\beta_{\text{super}} = 2- \phi = 2 - \gamma/D_p$, which implies   
\begin{equation}\label{eq_phi}
\phi = \frac{\gamma}{D_p}
\, .
\end{equation}

Inserting Eq.~(\ref{eq_phi}) in Eq.~(\ref{eq_pint_conec}) one can conclude that the two approaches are similar -- i.e.\ they lead to the same results -- when the Euclidean and social distances ($r$ and $d$, respectively) are related by $d \sim D_p \log_b(r)$
or 
\begin{equation}\label{eq_dr2}
r \sim b^{\frac{d}{D_p}}
\, .
\end{equation}
This result indicates that according to the models, Euclidean and social distances should be correlated, which is plausible since we live close to people we know \cite{Goldenberg2009}. 
Moreover, if Eq.~(\ref{eq_dr2}) holds, then the two approaches, in fact, represent the same urban system.

If we solve Eq.~(\ref{eq_dr2})  for the fractal dimension $D_p$, one gets
\begin{equation}\label{eq_arbes_dc}
D_p \sim \frac{d}{\log_b(r)}
\, .
\end{equation}
The network distance $d$ must be proportional to the logarithm of a ``mass'' since otherwise Eq.~(\ref{eq_arbes_dc}) would not comply with the definition of the fractal dimension \cite{BundeH1995Sec1}.
Interestingly, in the social-network model the number of nodes and the distance are related via $N(d) = b^d$, Eq.~(\ref{eq_net_Nd}), and consequently $d \sim \log_b(N)$; that is, $d$ is proportional to the logarithm of a ``mass''. 
Inserting in Eq.~(\ref{eq_arbes_dc}) yields $D_p \sim \log(N) / \log(r) $, which makes sense in terms of fractal geometry.

This means, if the fractal dimension $D_p$ relates population and space -- where population takes the role of mass and Euclidean distance the role of scale -- then the gravity models in both versions are equivalent.
This is remarkable since the various authors \cite{ribeirocity2017,yakubo2014,Arbesman2009}
developed their models independently employing different ideas and approaches.
Under this condition, the population is spatially located in fractal manner while at the same time it follows a hierarchical social network.
From Eq.~(\ref{eq_phi}) we conclude that the decay exponent $\gamma$ also relates to social ties, expressed by the parameter $\phi$, which gives one more insight about the $\gamma$ parameter and consequently how the interaction between people behave.

\subsubsection*{F.Ribeiro et al.\ model of sub-linear scaling}

The infrastructure scaling exponent $\beta_{\text{sub}}$ can also be deduced using the the gravity approach.
However, different from the Bettencourt model (Sec.~\ref{sec_bettecourt}), that focuses on the area to explain the sub-linear urban scaling, Ribeiro et al.\ \cite{ribeirocity2017} focus on the number of amenities in the city necessary to satisfy the people needs.
They use the idea that people tend to choose close places to buy products.

The model considers that the individual $i$ consumes $u_i$ quantities of a given \textit{individual need} product (e.g.\ bread) per period of time, and consequently the city, as a whole, consumes $U = \sum_{i=1}^N u_i$ units  
of this product during this period.
It is assumed that the demand is always fully provided by the city.
Suppose that the consumers can buy this product in $P$ amenities (e.g.\ bakeries, if the product is bread) distributed throughout the city, but these consumers choose, preferentially, the amenities that are close. Using the same idea discussed around Eq.~(\ref{eq_f_r}), the total supply (per time) by  the $k$-th amenity for the person $i$ will be given by
\begin{equation}\label{eq_fr}
f(r_{ik}) \propto \frac{1}{r_{ik}^{\gamma}} \, ,  
\end{equation}
where $r_{ik}$ is the Euclidean distance between them. 
Equation~(\ref{eq_fr}) can also be understood as the number of products the individual $i$ bought in the $k$-th amenity during a period of time. 
As $\gamma>0$, this person buys more products in closer amenities.    
The total demand of $i$ can then be computed by
\begin{equation}\label{Eq_potencial}
u_i = \sum_{k=1}^P f(r_{ik}) \, , 
\end{equation}
and consequently the total provision of the city is 
\begin{equation}
U \equiv \sum_{i=1}^N u_i = \sum_{i=1}^N \sum_{k=1}^P  f(r_{ik}) \, , 
\end{equation}
which can also be written as
\begin{equation}\label{Eq_U}
U = \sum_{k=1}^P \left( \sum_{i=1}^N  f(r_{ik})  \right) \, .
\end{equation}
If we consider that the population is homogeneously distributed in a fractal structure, as considered before (in Sec.~\ref{sec_gravity_simplest}), then the inner sum can be transformed into an integral  
that can be solved as before using
Eqs.~(\ref{eq_rho}),~(\ref{eq_dr}), and~(\ref{eq_fr}):
\begin{equation}
\sum_{i=1}^N  f(r_{ik})  \sim
\int f(\mathbf{r}) \rho(\mathbf{r}) d \mathbf{r} \sim
N^{1-\frac{\gamma}{D_p}} \, ,    
\end{equation}
which reveals that this sum, on average, is the same for all amenities. 
The right-rand side of this proportionality does not depend on the index $k$. 
Inserting this result in Eq.~(\ref{Eq_U}), the second sum transforms into a multiplication by the total number of amenities ($P$), leading to
\begin{equation}\label{Eq_U_scale}
U \sim P N^{1-\frac{\gamma}{D_p}}
\, .
\end{equation}

As we are dealing with individual need products, the empirical evidence suggests that the total consumption must \textit{scale linearly} with the population size, that is $U \sim N$. 
Under this condition, Eq.~(\ref{Eq_U_scale}) can be rewritten as
\begin{equation}\label{Eq_popt}
P \sim N^{\frac{\gamma}{D_p}},
\end{equation}
and therefore one gets the scaling exponent 
\begin{equation}\label{beta_infra}
\beta_{\text{sub}} = \frac{\gamma}{D_p},
\end{equation}
which governs the scaling properties of the number of amenities -- an infrastructure variable -- as a function of the population size. 
It is worth mentioning that together with Eq.~(\ref{eq_beta}) the rule of thumb $\beta_{\text{sub}}+\beta_{\text{sub}}=2$, Eq.~(\ref{eq_pint_res}), is fufilled.

In conclusion, this model
suggests that sub-linear exponent of the infrastructure occurs only if $\gamma < D_p$. 
This $\gamma$-range in fact characterizes the long-range interaction regime, which is a consequence of the city acting as an entire \textit{coupled system}, as mentioned in Sec.~\ref{sec_gravity_simplest} in the context of the gravity model and comparison to the Bettencout model.
In addition, the bigger the interaction range is, the bigger the economy of scales and consequently the lower are the infrastructure costs.

\subsubsection*{Findings and conclusions from gravity  models}

At least tree groups of authors independently employ gravity ideas to explain urban scaling -- here we show that they lead to consistent results.
This observation emphasizes the importance of the gravity approach to understanding urban phenomena, 
as it was already suggested qualitatively by Tobler with his first law of geography\footnote{``Everything is related to everything else, but near things are more related than distant things.''} \cite{Tobler1970}.

A novelty of the present work is the organisation of these ideas in a single and general framework that permits to identify similarities and common results. 
The various gravity models are equivalent in special cases, outside this overlap they represent variants. 
However, from consolidating the models we conclude that \emph{urban scaling is essentially a consequence of spatial distribution imposed by the geometry and the social ties that enhance or reduce interactions}.
Specifically, we make the following interpretations.
\begin{itemize}
\item \emph{Good access to all parts of the city.}
Increasing returns to scale require a geographically well connected city, allowing interactions within the entire city and permitting integrity of the city.
In practice, this can be achieved by an efficient transport system.
\item \emph{Influencers reaching distant parts of the city.}
The presence of outstanding influential people can integrate distant parts of the city and promote interconnectivity, resulting in a more pronounced urban scaling.

\item \emph{Interaction between socially distant people.}
The socio-economic scaling exponent is larger when socially distant people can interact better. We have demonstrated that under certain circumstances social and geographic distance are related.

\end{itemize}

\subsubsection{Molinero \& Thurner model -- infrastructure geometry}

This section presents the model proposed by Molinero and Thurner \cite{Molinero2021} which, as some of the models presented in the previous section, also employs geometrical considerations as primary factor responsible for urban scaling.
The authors introduce new ingredients to the discussion, like the city verticalization and the distinction between the fractal dimension of the population ($D_p$) and the fractal dimension of the infrastructure ($D_{\rm{infra}}$), i.e.\ the street network. 
As the street network rests on the two-dimensional earth surface, its dimension is constrained by $D_{\rm{infra}} \le 2$ \cite{Lu2016}.
Molinero \& Thurner argue that the distinction of population and infrastructure fractal dimensions is essential to the scaling laws observed across cities. 

The population is located in houses and buildings, which are situated along the streets. 
The authors argue, that if we neglect the vertical extent, then the fractal dimension of the population would be very similar to the fractal dimension of the street structure.
This is similar to what the previous models consider, i.e.\ $D_{\rm{infra}} = D_p$. 
However, as Molenero and Thurner argue, the cities have a vertical component that cannot be disregarded, which constrains $D_p$ to the interval $D_{\rm{infra}} \le D_p \le D_{\rm{infra}} + D_h$,  where $D_h$ is the dimension associated with the city building height.
If one considers that people fulfill all three-dimensional embedded components of the infrastructure, then \begin{equation}
D_p = D_{\rm{infra}} + D_h 
\, .   
\end{equation}

The population fractal dimension $D_p$ is defined by the power-law relation between population and a linear metric $r$, that is 
\begin{equation}\label{eq_def_Dp}
N \sim  r^{D_p}
\, .
\end{equation}
In the same way, the street network fractal dimension $D_{\rm{infra}}$ is defined by the power-law relation between the
\emph{street network total length} $L_{\rm{tot}}$ and a linear metric
\begin{equation}\label{eq_L_total}
L_{\rm{tot}} \sim  r^{D_{\rm{infra}}} \, .
\end{equation}

In addition, as proposed by the authors, the number of individuals can be written using these two dimensions 
\begin{equation}\label{eq_Nc}
N \sim C_c \cdot r^{D_p}
\end{equation}
and 
\begin{equation}\label{eq_C_net}
N \sim C_{\rm{infra}} \cdot r^{D_{\rm{infra}}}
\, .
\end{equation}
In Eq.~(\ref{eq_Nc}), $C_c$ can be understood as the number of people living in a \emph{cube} of size~1 (in any units). 
This means $r^{D_p}$ is the number of non-empty cubes of size~1 in the city. 
Such a cube can be, for instance, a house, an apartment, or a floor. 
Theoretically, $C_c$ must be scale-invariant, that is, $C_c \sim N^0$, because of the physical limit to accommodate a maximum number of people in a house/apartment/floor. 
The authors verified that $C_c$ increases with the city size for cities smaller than $100.000$ inhabitants, but it indeed saturates and stabilises for cities larger than that. 
Similarly, the other quantity, $C_{\rm{infra}}$ in Eq.~(\ref{eq_C_net}), can be understood as the number of people living in a \emph{square} of size~1. 
This means $r^{D_{\rm{infra}}}$ is related to the number of non-empty squares of size~1. 
Indeed $C_{\rm{infra}}$ represents the projection of the three-dimension space population into a two-dimensional plane (vertical projection).
In contrast to $C_c$ that is constant for a sufficiently large population, $C_{\rm{infra}}$ grows with the population size obeying a power-law relation $C_{\rm{infra}} \sim N^{0.09}$ in the UK and the authors observe similar results in other countries. 
It reveals an absence of a typical size value $C_{\rm{infra}}$.  

The relation between $C_{\rm{infra}}$ and $C_c$ can be obtained equalling Eqs.~(\ref{eq_Nc}) and~(\ref{eq_C_net}) to get
\begin{equation}\label{eq_cnet}
C_{\rm{infra}} = C_{c} r^{D_p - D_{\rm{infra}}}.  
\end{equation}
Using Eq.~(\ref{eq_def_Dp}) in this relation and considering that $C_c \sim N^0$, leads to
\begin{equation}
C_{\rm{infra}} \sim N^{1-\frac{D_{\rm{infra}}}{D_P}}
\, .
\end{equation}

The saturation of $C_c$ for sufficiently large cities and the power-law relation between $C_{\rm{infra}}$ and $N$ implies that the densification of the ``cube'' of size~1 happens until the city population reaches a limit (around $N=100.0000$ inhabitants). 
For cities larger than this, the number of people in this cube is stabilized, but to increase the number of people per square meter (that is, to continue increasing $C_{\rm{infra}}$ with $N$), the city starts to grow vertically. 

Defining such quantities, one can derive the urban scaling exponents. 
Using Eqs.~(\ref{eq_L_total}) and~(\ref{eq_Nc}) one can show that $L_{\rm{tot}} \sim N^{\frac{D_{\rm{infra}}}{D_p}}$, i.e.\

\begin{equation}\label{Eq_molinero_beta_sub}
\beta_{\text{sub}} =  \frac{D_{\rm{infra}}}{D_p} \, .
\end{equation}
This result implies that the urban scaling exponent is the result of the relationship between the two fractal structures, namely population and infrastructure (street) network.
In addition, the emergence of non-linearity ($\beta \ne 1 $) happens because of the difference of the fractal dimensions of these structures, and the sub-linearity ($\beta < 1$) is due to $D_{\rm{infra}} \le  D_p$.  

The authors also consider that the socio-economic variable must be dependent on the number of interactions in the city. 
To estimate this number they consider, by hypothesis, that  the number of interactions inside a square of size~1 will be proportional to the maximum number of interaction $C_{\rm{infra}} (C_{\rm{infra}}-1)/2 \sim C_{\rm{infra}}^2$.
With this consideration, the total number of interactions in the city ($N_{\text{int}}$) is given by the number of interactions inside a square of size~1 multiplied by the number of squares of this size \begin{equation}\label{eq_Nint}
N_{\text{int}} \sim C_{\rm{infra}}^2 r^{D_{\rm{infra}}} \, .
\end{equation}
Using Eqs.~(\ref{eq_cnet}) and~(\ref{eq_Nc}) one can show that $N_{\text{int}}~\sim~N^{2- \frac{D_{\rm{infra}}}{D_p}}$, and therefore, considering that $Y \sim N_{\text{int}}$, 
\begin{equation}\label{eq:mtbeta}
\beta_{\text{super}} = 2- \frac{D_{\rm{infra}}}{D_p} \, .
\end{equation}
It shows, which role fractal structures (population and streets networks) play in urban scaling. 
Urban scaling emerges as a result of an imbalance of where people live and the structure on which they move.
It is also important to stress that, as demonstrated by the authors, while the population and infrastructure fractal dimensions ($D_P$ and $D_{\text{infra}}$) vary largely for the individual cities, the empirical ratio $D_{\text{infra}}/D_P$ is remarkably robust (for thousands of cities) and around $D_{\text{infra}}/D_P \approx 0.86$ \cite[Fig.~2]{Molinero2021}.
The closeness of this numeric value to the empirical value of the infrastructure scaling exponent ($\beta_{\text{sub}} \approx 0.85$) is remarkable and puts this theory (conform Eq.~(\ref{Eq_molinero_beta_sub})) as one of the most successful in explaining urban scaling in the context of infrastructure.

\subsubsection*{Conection with the gravity models}

Apparently Eqs.~(\ref{eq_beta}), $\beta_{\text{super}} = 2- \frac{\gamma}{D_p}$, and Eq.~(\ref{eq:mtbeta}), $\beta_{\text{super}} = 2- \frac{D_{\rm{infra}}}{D_p}$, have a very similar form and only differ in the numerator, implying $\gamma=D_{\rm{infra}}$.
Indeed, comparing these two independent results provides an additional interpretation of the gravity exponent $\gamma$ and suggests a more fundamental explanation of the Molinero and Thurner result.

In the gravity model context -- Section~(\ref{sec_gravity}) -- the super-linear scaling exponent in Eq.~(\ref{eq:mtbeta}) emerges when the probability of interaction between two individuals separated by the Euclidean distance $r$ is given by
\begin{equation}\label{eq_pint_molinero}
p_{\rm{int}}(r) \sim \frac{1}{r^{D_{\rm{infra}}}} \, ,
\end{equation}
where we have replaced $\gamma$ with $D_{\rm{infra}}$ in Eq.~(\ref{eq_f_r}).
This result indicates how space -- or the street network structure, here represented by its fractal dimension $D_{\rm{infra}}$, -- affects the connection between the people.
If Eq.~(\ref{eq_pint_molinero}) holds, then the more compact the street network is (larger $D_{\rm{infra}}$), the smaller is the interaction range. 
In other words, with increasing density fewer parts of the city can be access by the individuals. 
Remains to interpret what $r^{D_{\rm infra}}$ means or what quantity it represents. 
There are at least two candidate quantities that scale as $ r^{D_{\rm infra}}$, and that could be responsible for this impedance on the individuals' interaction. 
First, the \textit{total length} inside a circle of radius $r$, which according to the definition Eq.~(\ref{eq_L_total}) scales as $\sim r^{D_{\rm infra}}$. 
Second, the \textit{number of sites/places/houses} inside a circle of radius $r$; as the houses are coupled to the streets, this number must also scale as $r^{D_{\rm infra}}$).  
In this sense, Eq.~(\ref{eq_pint_molinero}) suggests a quantitative way to understand how the structure of the streets affects the interaction of the people and, consequently, how it 
reverberates on urban scaling.

\subsection{Gomez-Lievano et al. -- model of required factors}
\label{sec:gomezlmodel}

This section presents the Gomez-Lievano et al.\ 2016 model of required factors \cite{Gomez-Lievano2016}, which belongs to the intra-city model category, but differs from the other models presented so far. 
While other intra-city models are based on \textit{human interaction} premises, this one is rather based on required factors within the city.
Consequently, the general framework introduced in Sec.~(\ref{sec:huminter}) does not apply here.
The model employs concepts of economic complexity and cultural evolution to explain the origin of urban scaling.
The main idea of the model is that an urban socio-economic phenomenon occurs when a number of necessary complementary factors are available in the city.

Consider $M$ as the number of possible factors required for a particular socio-economic activity.
For instance, if $Y$ is the total number of patents in a city, $M$ can be the number of different skills and capabilities needed by an individual to develop a patent.
That is, $M$ is a measure of the ``sophistication'' (complexity, difficulty, etc.) of the phenomenon in question \cite{Gomez-Lievano2018}.
Suppose that each one of these factors is provided by the respective city with probability $z$.
Then, if these factors are independent, the probability that the city provides $m$ of these $M$ factors, say $p(m)$, will follow the binomial distribution 
\begin{equation}\label{Eq_dist_m}
p(m) = \frac{M!}{m! (M-m)!}  z^m (1-z)^{M-m}
\, .
\end{equation}
According to the authors, the parameter $z$ can be interpreted as a measure of urban diversity. 

One particular individual $i$ will only succeed to implement/develop this socio-economic activity if she/he only needs factors that the city can provide.
The authors introduce the binary random variable $w_i$ that 
is $w_i=1$ when the individual succeeds or $w_i=0$ when the individual fails to implement this socio-economic activity.
An analogous proposition, but without scaling analysis, was studied in \cite{Hausmann2011}.

Let us call $p(w_i =1| m)$ the probability of success given that the city 
provides $m$ factors.
This probability is identical to the probability that this individual \textit{does not need} the $(M-m)$ factors that the city \textit{does not provide}.
If we denote $q$ as the probability that $i$ needs any given factor -- i.e.\ it is a measure of the ability of the individuals \cite{Gomez-Lievano2018} -- then we can write
\begin{equation}\label{Eq_dist_x}
p(w_i =1| m)= (1-q)^{M-m} \, . 
\end{equation}

The total socio-economic activity in the city will be given by the expected value of the aggregate output $Y = \sum_{i=1}^N w_i$, that is 
\begin{equation}\label{eq_Yglx}
\langle Y \rangle = N \sum_{m=0}^M  p(w_i =1, m)
\, ,
\end{equation}
where $p(w_i =1, m)$ is the joint probability also given by   
\begin{equation}
p(w_i =1, m)  = p(w_i =1| m) p(m) \, .
\end{equation}
Then, using the probability distributions Eqs.~(\ref{Eq_dist_m}) and~(\ref{Eq_dist_x}), the expected value of the totality of this socio-economic activity (from Eq.~(\ref{eq_Yglx}) and using binomial properties) will be
\begin{equation}\label{Eq_Ygl1}
\langle Y \rangle  =  N \Big(1- q(1-z) \Big)^M  
\, .  
\end{equation}
The term $q(1-z)$ represents the probability that a factor is neither possessed by the individual nor by the city, which is likely to be very low. 
This can happen for sufficiently small $q$, e.g.\ if the number of skills per individual is large; but also for $z \to 1$, e.g.\ cities tend to be very diverse places, in the sense that there are very few factors that cannot be hired/found/bought therein. 
Then, for $q(1-z)$ sufficiently low Eq.~(\ref{Eq_Ygl1}) yields
\begin{equation}\label{Eq_Ygl}
\langle Y \rangle \approx  N e^{-Mq(1-z)} 
\, .  
\end{equation}

As argued by the authors, this expression corresponds to a power-law if $z$ is a logarithmic function of the population size,
\begin{equation}\label{Eq_rN}
z(N) = a + b \ln(N)
\, ,
\end{equation}
where $a$ and $b$ are constants.
It turns Eq.~(\ref{Eq_Ygl}) into the power-law $\langle Y \rangle = Y_0 N^{\beta_{\text{super}}}$, where 
\begin{equation}\label{eq_gomez-beta}
\beta_{\text{super}} =  1 + Mbq 
\, .     
\end{equation}

The authors argue that Eq.~(\ref{Eq_rN})
can be explained by considering the way cities accumulate factors as they grow in size. 
According to them, this relationship emerges if factors are added to cities as they increase their size and a selection process occurs in which only the best or most useful factors survive. 
Such cumulative evolutionary processes have been analyzed in the cultural evolution literature \cite{Henrich2004}, and they give rise to factors accumulating with the logarithm of population size.

To sum up, the model predicts that the super-linearity of the urban scaling exponent is due to
(i) the number of factors a given socio-economic activity requires to happen (expressed by $M$),
(ii) the capacity of the city to provide the necessary complementary factors ($b$); and 
(iii) the dependence of individuals to get factors from their urban environment ($q$).
The larger these quantities are, the more pronounced is the super-linearity.

\subsection{Gomez-Lievano et al.\ -- extreme value model}

Gomez-Lievano et al.\ 2021 \cite{Gomez-Lievano2021} argue that the super-linear scaling may not be a consequence of increasing returns to scale, as it is usually assumed. 
In this work, the authors propose a hypothetical situation where the non-linearity of the urban scaling could emerge even without an interaction process between the agents.
They show that non-linearity can emerge by a selection process acting on independent random variables.
In this sense, urban scaling would rather represent an artefact.

To demonstrate this argument, the authors propose the following model.
Assume that a given individual has a productivity $w$, which is an independent random variable, and consequently, it does not depend on the size of the city he/she lives in. 
Moreover, this productivity is log-normally distributed, following
\begin{equation}
p_w(w | x_0, \sigma^2) = \frac{1}{w \sqrt{2\pi \sigma^2}} e^{- \frac{(\ln w - \ln w_0)^2}{2\sigma^2}}
\, , 
\end{equation}
where $w_0$ and $\sigma$ are positive parameters, such that $\ln w_0 = \langle \ln w \rangle$ is the expectation value of $\ln w$, $\sigma^2 = {\rm VAR}[\ln w]$ is the variance, and the productivity expectation value is given by $\mu = \langle w \rangle = w_0 e^{\frac{\sigma^2}{2}}$.
The choice of a log-normal probability density function (pdf) is justified by some empirical evidence suggesting that productivity across workers, measured indirectly by wages, follows such distribution \cite{Eeckhout2014,Combes2012}. 

The authors model the total socio-economic production of the $c$-th city as 
the sum of the productivity of all citizens of the city, that is 

\begin{equation}
Y_c (N_c) = \sum_{i=1}^{N_c}w_i^c    
\end{equation}
where $N_c$ is the population of the $c$-th city.
The expectation value of this production can be computed as $\langle Y(N) \rangle = \sum_{i=1}^N \langle w_i \rangle$ which implies $\langle Y(N) \rangle = N  \langle w_i \rangle $.  
Similarly, for any other city with population $\lambda N$, where $\lambda$ is an increase factor, one gets $\langle Y(\lambda N) \rangle = \lambda N \langle w_i \rangle $. 
Consequently, it is possible to identify
\begin{equation}
\frac{  \langle Y(\lambda N) \rangle  }{\lambda N}  =     
\frac{  \langle Y(N) \rangle  }{ N} \, ,
\end{equation}
which means the per-capita socio-economic output remains unchanged in the face of an increase in population, which shows that a basic situation of independent random variables yields $\beta = 1$ (linear scaling).

However, scaling properties can appear if one considers the production $Y$ as proportional to the maximum value of the productivity in the city, i.e.\
\begin{equation}\label{eq_Y_glmax}
Y(N) \equiv \rm{max}\{  w_1, \cdots, w_N\} \, . 
\end{equation}
This represents the case where the productivity is dominated by the most productive individuals.

For convenience, lets rewrite $w_i$ as $w_i = e^{\sigma z_i + \ln w_0}$, where $z_i$ is an independent random variable sampled from a normal distribution with mean~0 and variance~1. 
The parameter $w_0$ is choose to be $\ln w_0 = - \frac{\sigma^2}{2}$, by convenience only, to promotes  $\langle w_i \rangle = e^{ \ln w_0 + \frac{\sigma^2}{2  }} = 1$.
Then Eq.~(\ref{eq_Y_glmax}) can be rewritten as 
\begin{equation}\label{eq_Y_gl_new}
    Y(N) \sim   e^{ \sigma Z(N)  - \frac{\sigma^2}{2  }}
\end{equation}
for $\sigma \gg \sqrt{\ln N}$. 
Here $Z(N) \equiv \rm{max} \{  z_1, \cdots, z_N\}$, and it is well known that it 
converges to the \textit{Gumbel distribution} \cite{Coles2001, Leadbetter1983},
leading to
\begin{equation}
    Z(N) \sim \sqrt{2 \ln N} \, .
\end{equation}

Hence, from Eq.~(\ref{eq_Y_gl_new}) one obtains
\begin{equation}\label{eq_Y_glnew2}
    Y(N) \sim   e^{  \sigma \sqrt{2 \ln N}   - \frac{\sigma^2}{2}}.
\end{equation}
Considering the power law of the form $Y\sim N^\beta$ holds, the scaling exponent is given by 
\begin{equation}
\beta = \frac{\frac{d}{dN} \ln Y}{ \frac{d}{dN} \ln N  } \, .
\end{equation} 
Applying such derivative to Eq.~(\ref{eq_Y_glnew2}) yields
\begin{equation}
    \beta = \frac{\sigma}{  \sqrt{2 \ln N} } \, ,
\end{equation}
which is valid for $N \ll \sigma$, implying super-linearity.
Otherwise prevails $\beta = 1$, as discussed previously.

The result of this model has at least two consequences.
First, increasing super-linear scaling, and increasing returns to scale, can be a mere artefact because it can happen from the selection process of independent random variables.
Second, it allows different regimes, linear and super-linear, depending on the relation between the size of the city and the variance in the distribution of workers' productivity. 
Small cities ($N \ll \sigma$) would exhibit super-linear scaling, while larger cities ($N \gg \sigma$) would exhibit linear regime.

\section{Inter-city Models}\label{sec:intercmodels}

The models presented so far are based on city internal (intra-city) aspects.
However, cities are not closed or isolated objects and, indeed, cities are in constant interaction with each other, be it by relations among individuals/firms from different cities \cite{Alderson2010}, be it by the very migration flows between cities \cite{Curiel2018}. 
Consequently, processes taking place between cities must, in some way, interfere with the productivity and the use of infrastructure. 
It is natural to also elaborate how such interactions between cities, that is \textit{inter-city} processes, could explain urban scaling or interfere with it, even if only in a second-order approximation.
In this section models that can be assigned to the intra-city category are presented. 
This line of research is less advanced, as suggested by a smaller number of models and a wider range of concepts, and some of them only comprise of a qualitative approach.

\subsection{Pumain et al.\ model of technological diffusion}

Pumain et al.\ \cite{pumain2006} argue that non-linearities are due to interactions within the \emph{system} of cities. 
More specifically, they propose that non-linearity emerges through a hierarchical diffusion process of innovations, from the largest to the smaller cities. 
According to this proposition, the innovation process is disproportionately higher in larger cities.
Super-linear scaling of economic indicators represents a stage of emergence of new technologies, which take places in larger cities; linear scaling represents the diffusion stage, from larger cities to towns and small cities of the system, and sub-linear scaling represents the mature stage of technologies, also characterised by decay or substitution processes.
The merit of this theory is that it brings the interconnection between cities to the forefront, evidenced, for instance, by Zipf's law which reveals some kind of hierarchical structure in urban systems \cite{auerbacghesetz1913,zipfhuman2012,berrycity2012,RybskiD2013,nitschzipf2005,soo2005zipf,cottineaumetamipf2017,RozenfeldRGM2011,gabaix1999}.

\subsection{Gomez-Lievano et al.\ relation}

Gomez-Lievano et al. \cite{Gomez-Lievano2012}  propose a statistical framework to characterize urban scaling and city size distributions.
In simple terms, they derive the scenario under which the city population sizes $N$  follow Zipf's law
\cite{gabaix1999,Berry2012,BARTHELEMYbook}, i.e.\ a power-law distribution according to
\begin{equation}
P(N) \sim N^{-\alpha}
\label{eq:PNalpha}
\, ,
\end{equation}
with $\alpha \approx 1$. Here,  $P(N)$ represents the \textit{(complementary) cumulative distribution function} (ccdf).
In addition, the authors consider, based on empirical evidences, that the 
ccdf of a given urban indicator $Y$, namely $P(Y)$, also follows a power-law
\begin{equation}
P(Y) \sim Y^{-\alpha_Y} \, ,
\label{eq:PYalpha}
\end{equation}
where $\alpha_Y$ is usually different from 1 for socio-economic and infrastructure urban variables. 

While the authors use a probabilistic characterization of urban scaling, in the following we present a back-of-an-envelop derivation.
If $p(N)$ is the \textit{probability density function} (pdf), in the sense that $P(N) = \int p(N)dN$, and similarly for $p(Y)$, then one distribution can be transformed into the other obeying the density transformation 
\begin{equation}
p(N)dN=p(Y)dY
\, .
\end{equation}
If we use the distributions Eq.~(\ref{eq:PNalpha}) and~(\ref{eq:PYalpha}) then we can write the integrals
\begin{equation}
\int N^{-\alpha-1}dN \sim \int Y^{-\alpha_Y-1}dY 
\, ,
\end{equation}
which leads to $N^{-\alpha} \sim Y^{-\alpha_Y}$ and 
\begin{equation}
Y \sim N^{\frac{\alpha}{\alpha_Y}}
\, .
\end{equation}
Finally, comparison with Eq.~(\ref{eq_Ysuper}) provides

\begin{equation}\label{eq:gleq23}
\beta=\frac{\alpha}{\alpha_Y}
\, .   
\end{equation}
This means that the scaling exponent is directly related to the Zipf exponent and vice-versa; Zipf's law and urban scaling are connected phenomena.
Recently, various relations were proposed explaining the connection between the Zipf and scaling exponents quantitatively \cite{HRibeiro2021} and qualitatively  \cite{hypotheses2021}. 
More specifically, urban scaling transforms the Zipf distribution ($\alpha\approx 1$) into another power-law distribution with another exponent $\alpha_Y$ that differs from $\alpha$ if $\beta\ne 1$.

However, Ribeiro et al.\ \cite{HRibeiro2021} argue that Eq.~(\ref{eq:gleq23}) only represents an upper limit.
By permuting the values $N$ and $Y$ from different cities, the association is destroyed so that correlations vanish ($\beta\approx 0$) -- but the distributions and the exponents $\alpha$, $\alpha_Y$ remain unaffected. 
This thought experiment leads to a situation where Eq.~(\ref{eq:gleq23}) is violated.
However, different degrees of correlations permit $\beta$-values up to $\frac{\alpha}{\alpha_Y}$.
In other words, Eq.~(\ref{eq:PNalpha}) and $Y\sim N^\beta$ imply Eq.~(\ref{eq:PYalpha}) but Eqs.~(\ref{eq:PNalpha}) and~(\ref{eq:PYalpha}) do not imply $Y\sim N^\beta$ with $\beta=\frac{\alpha}{\alpha_Y}$.

\subsection{H.Ribeiro et al.\ model -- country scaling}

Ribeiro et al.\ \cite{HRibeiro2021} empirically relate Zipf's law for cities and urban scaling.
Based on data for many countries, they find correlations between the Zipf-exponent $\alpha$ and the urban scaling exponent $\beta$ for GDP.
In order to explain these correlations, the authors argue that for a given total urban population the country-wide urban GDP is fixed and different values of $\alpha$ require an adjustment of $\beta$ so that the country-wide aggregate is preserved.
The same argument is used vice versa, i.e.\ the authors make no statement about the direction of a possible causality.

\begin{figure}
	\begin{center}
\includegraphics[width=\columnwidth]{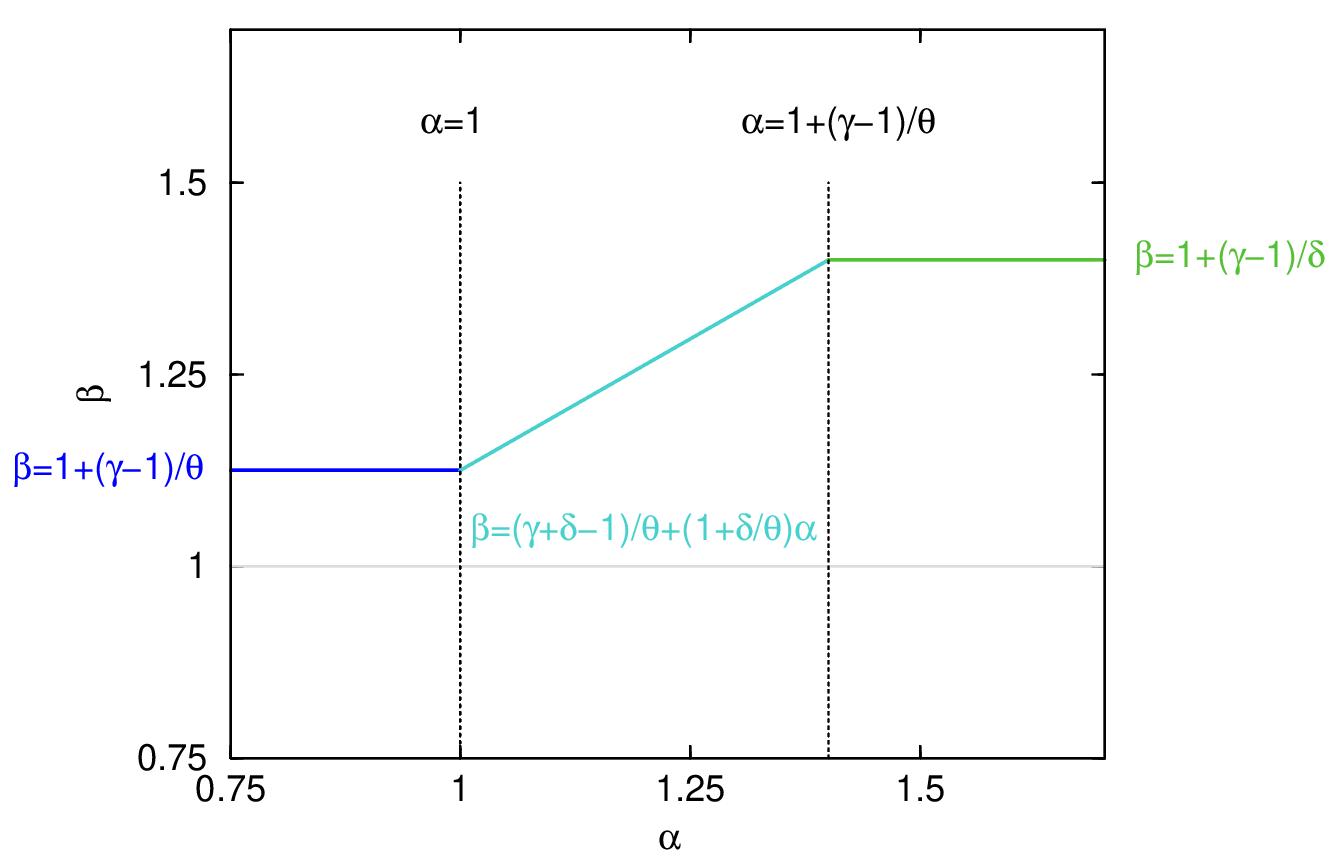}
\end{center}
	\caption{ \label{Fig_haroldo} 
Illustration of Eq.~(\ref{eq:hrbafinal}) relating the urban scaling exponent $\beta$ and the Zipf exponent $\alpha$.
Here the parameters $\gamma=1.1$, $\delta=0.25$, and $\theta=0.8$ have been used.
Source: adapted after \cite{HRibeiro2021}.
	}
\end{figure}

Combining Zipf's law and urban scaling, the total contry-wide output of a considered (additive) socio-economic urban metric is given by
\begin{equation}
Y^*=\sum_{N=N_{\text{min}}}^{N_{\text{max}}}Y(N)h(N) \sim \int_{N=N_{\text{min}}}^{N_{\text{max}}}N^\beta h_1 N^{-\alpha-1} dN
\, .
\label{eq:HarYsum}
\end{equation}
where $h(N)$ is the frequency function, which according to Zipf's law Eq.~(\ref{eq:PNalpha}) is 
$h(N) = h_1 N^{-\alpha -1}$; 
and $h_1$ is the normalization constant from the city size distribution. 
The constant $h_1$ can be obtained from the total urban population
\begin{equation}
N^*=\sum_{N=N_{\text{min}}}^{N_{\text{max}}} Nh(N) \approx \int_{N_{\text{min}}}^{N_{\text{max}}}N h_1 N^{-\alpha-1} dN
\, .
\end{equation}
These equations are solved by considering that the sizes of the largest and smallest cities in a country depend on the total population of that country, following power-laws,
\begin{equation}\label{eq:NmaxNmin}
N_{\text{max}} = b (N^*)^\theta \qquad \textrm{and} \qquad 
N_{\text{min}} = a (N^*)^\delta
\, ,
\end{equation}
where $a,b,\theta,\delta$ are constants.
Introducing these relations in Eq.~(\ref{eq:HarYsum}), the authors obtain
\begin{equation}
        Y^* \sim 
        \frac{(\alpha-1) N^*}{\alpha-\beta}
        \left( 
        \frac{a^\beta b^{\alpha} (N^*)^{\delta\beta + \theta\alpha} - a^{\alpha} b^\beta (N^*)^{\theta\beta+\delta\alpha}}
             {a b^{\alpha} (N^*)^{\delta +\theta\alpha} - a^{\alpha} b (N^*)^{\theta + \delta \alpha}}
        \right) \, .
\label{eq:Har4trms}
\end{equation}
Country populations are large ($N^*\gg 1$) and in this limit the total aggregate urban metric becomes
\begin{equation}
Y^* = Y_0^*(N^*)^\gamma
\label{eq:Hargamma}
\, ,
\end{equation}
where $Y_0^*$ and $\gamma$ are constant.
Which of the four terms in the parenthesis of Eq.~(\ref{eq:Har4trms}) dominates in the limit, depends on the values of the exponents $\alpha,\beta,\delta,\theta$.
Considering Eq.~(\ref{eq:Har4trms}) in the limit $N^*\gg 1$ and comparing with Eq.~(\ref{eq:Hargamma}), the exponents can be solved for $\beta$ yielding
\begin{equation}
  \beta = \begin{cases}
  1 + \frac{\gamma-1}{\theta} & 0<\alpha\leq 1\\
  \frac{\gamma+\delta-1}{\theta} + \left(1-\frac{\delta}{\theta}\right)\alpha & 1<\alpha<1+\frac{\gamma-1}{\delta}\\
  1 + \frac{\gamma-1}{\delta} & \alpha\geq 1+\frac{\gamma-1}{\delta}
  \end{cases} \, ,
\label{eq:hrbafinal}
\end{equation}
for $\gamma>1$ and $\delta<\theta$ (the authors provide similar expressions for other conditions).
It is a step-wise function (see Fig.~\ref{Fig_haroldo}), which in the middle regime exhibits a linear relation between $\beta$ and $\alpha$.
Solving for $\beta$ does not mean that there is a causality from $\alpha$ on $\beta$.

Last, it needs to be mentioned that $\alpha$ and $\beta$ are exponents across cities and there is one value each for a country.
The other exponents $\gamma,\delta,\theta$ are country scaling exponents across countries. 
This means, considering a country or a similar confined region as it is generally done in this paper, the country scaling exponents are constants.
E.g.\ for a given country population $N^*$ Eq.~(\ref{eq:Hargamma}) with $\gamma$ implies a fixed country aggregate $Y^*$.

Comparison of Eq.~(\ref{eq:gleq23}) and the middle regime of Eq.~(\ref{eq:hrbafinal}) leads to $\frac{\gamma+\delta-1}{\theta}=0$ and $(1-\frac{\delta}{\theta})=\frac{1}{\alpha_Y}$.
It supports the compatibility of both views.
Fitting the country scaling relationships Eqs.~(\ref{eq:NmaxNmin}) and (\ref{eq:Hargamma}), Ribeiro et al.\ \cite{HRibeiro2021} report $\gamma\approx 1.31$, $\delta\approx 0.20$, and $\theta\approx 0.79$, approximately confirming the first relation and leading to $\alpha_Y\approx 1.34$.
However, as the country scaling relations only describe the mean field, the value $\alpha_Y\approx 1.34$ needs to be interpreted in the same sense and one needs to keep in mind some spread around it.

\subsection{Altmann et al. model -- attractiveness token}

The last model to be discussed is the one developed by Altmann~\cite{Altmann2020} based on an approach initially proposed by Leit\~ao et al.\ \cite{leitao2016}.
This model differs from the other models presented here in the sense that it does not lead to the scaling exponent as an emergent phenomenon. 
However, it is essential to mention it because the model builds on interactions between individuals of different cities, introducing ideas about integrating intra- and inter-city aspects. 
Moreover, this model suggests ideas that could be incorporated into the intra-city models, in order to expand them to an inter-city approach. 

The model considers an urban system composed of $N_{\rm{cit}}$ cities, where $N_c$ is the population of the $c$-th city, in such a way that the total population of the system is $ N^* = \sum_{c=1}^{N_{\rm{cit}}} N_c$.
There are also $Y^*$ tokens that are randomly assigned to the people of the system. 
This token can be, for example, a patent or an socio-economic output.
Let's denote $p(i)$ the probability that one token is attributed to the individual $i$ out of $N^*$ individuals. 
Altmann proposes that this distribution is a function of the attractiveness $x_i$ of this individual \cite{Altmann2020}. 
The attractiveness would be related to the ability of the individual to attract one of the tokens -- for example, charisma, leadership, or professional training. 
In other words, token attractiveness can also be understood in an active sense, i.e.\ related to the skills of an individual or to which extent an individual is able to convince others. 
This quantity can be thought of as the result of the connection between this individual and all the others, in the form

\begin{equation}\label{eq_ass_alter}
x_i \propto  \sum_{j=1}^{N^*} p_{\rm{int}}(r_{ij})
\end{equation}
where $p_{\rm{int}}(r_{ij})$ is a measure of the interaction between the individuals $i$ and $j$, that are at the distance $r_{ij}$ one each other. 
Here $r_{ij}$ can be within or across the boundaries of a city, since $i$ and $j$ can belong to the same or to different cities.

Note that the attractiveness $x_i$ also appears in the Yakubo et al.\ model (see Sec.~\ref{sec_yakubo}), and $p_{\rm{int}}(r_{ij})$ is the probability of interaction in the context of gravity models (see Sec.~\ref{sec_gravity}).
This way, $p_{\rm{int}}(r_{ij})$ can, for instance, obey the power-law form given by Eq.~(\ref{eq_f_r}). Moreover, Eq.~(\ref{eq_ass_alter}) can also recover a strict intra-city process when $p_{\rm{int}}(r_{ij}) = 0$ for $i$ and $j$ residing in different cities. 
There is a subtle but essential difference between the attractiveness considered by Yakubo et al.\ and the one considered by Altmann.  In the former, the attractiveness \textit{implies}  the probability of interaction (as given by Eq~(\ref{eq_pint2_yak})); while in the latter the attractiveness is a \textit{consequence} of the probability of interaction (as given by Eq.~(\ref{eq_ass_alter})).

Returning to the model, the probability that a token is allocated to a particular city $c$ is $p_{\rm{tok}}(c) = \sum_{i \in c} p(i)$, and the expected number of tokens at this city
can be computed using this probability, i.e.\
\begin{equation}\label{eq_Yk_tokens}
Y_c = Y^* \cdot p_{\rm{tok}}(c)
\, . 
\end{equation}

Altmann considers, by hypothesis, a non-linear efficiency of individuals, expressed by
\begin{equation}\label{Eq_pi}
p(i) = \frac{x_i^{1- \beta}}{Z(\beta)}
\, ,   
\end{equation}
where  $Z(\beta)$ is the normalization constant given by the sum $\sum_{i=1}^{N^*} p(i) =1$. 
In the particular case that $x_i$ is the same for all individuals of the same city $c$, that is $x_i = x_c$ for any $i$ belong to city $c$, the probability of this city to attract a token is

\begin{equation}
p_{\rm{tok}}(c) = N_c \cdot \frac{x_c^{1-\beta}}{Z(\beta)}.    
\end{equation}
Then, from Eq.~(\ref{eq_Yk_tokens}), 
the expected number of tokens at the $c$-th city is given by 
\begin{equation}
Y_c = Y^* \cdot N_c \cdot \frac{x_c^{1-\beta}}{Z(\beta)} 
\, .  
\end{equation}

The probability to observe the set $\{ Y_c \}_{c=1,\cdots,N_{\rm{cit}}}$ in the set of cities of size $\{ N_c \}_{c=1,\cdots,N_{\rm{cit}}}$ is indeed a multinomial distribution 
\begin{widetext}
\begin{equation}\label{eq_likelihood}
P(Y_1, \cdots, Y_{N_{\rm{cit}}} |  N_1, \cdots, N_{N_{\rm{cit}}} ) = \frac{Y^*!}{  \prod_{c=1}^{N_{\rm{cit}}} Y_c !  }    \prod_{c=1}^{N_{\rm{cit}}} \left( \frac{ N_c \cdot x_c^{\beta-1}}{Z(\beta)}       \right)^{Y_c}
\, .
\end{equation}
\end{widetext}
It correspond to the likelihood of the data $\{ Y_c \}_{c=1,\cdots,N_{\rm{cit}}}$ given a fixed population  $\{ N_c \}_{c=1,\cdots,N_{\rm{cit}}}$. 

The first version of the model \cite{leitao2016} is without spatial interactions. 
However, in the version presented in \cite{Altmann2020}, Altmann has generalized the model to take into account spatial interactions between individuals by Eq.~(\ref{eq_ass_alter}). 
He investigates the parameter $\beta$ that maximizes the posterior of this likelihood, varying the \textit{typical interaction distance} $r_0$ that governs the interaction range between the individuals. 
For instance, in the case where the probability of interaction is governed by an exponential decay $p_{\rm{int}}(r_{ij}) \sim e^{- \ln(2) r_{ij}/r_0}$, implies that the interaction becomes 0.5 for $r_{ij} = r_0$ and 1 for $r_{ij}=0$ (intra-city case).

He finds that the value of $\beta$ that maximizes Eq.~(\ref{eq_likelihood}), say $\beta^*$, is a function of the typical distance $r_0$, i.e.\ $\beta^* = \beta^*(r_0)$.
In addition, he finds that $\beta^*$ has a maximum value for $r_0 \ne 0$, which reflects an effect from an inter-city process.
For instance, for the urban GDP of Brazilian cities, $\beta^*$ is maximized when $r_0 = 14.6$\,km, suggesting that this is the typical interaction range for this system of cities.  

Altmann argues that urban scaling is a consequence of the non-linearity of the individuals' efficiency that depends on the size of the city they live in, modelled by Eq.~(\ref{Eq_pi}), which is in line with what other authors argue \cite{bettencourt2013}. 
Another distinguishing aspect of Altmann's work is a systematic way -- via the likelihood~(\ref{eq_likelihood}) -- to test different models, with their parameters being extracted and tested from data. 
It involves, according to Altmann, separating the analysis in two steps. 
First ``explaining'' the emergence of scaling (the model) and second, making a fit to the data (by maximum likelihood approach).

For future work, it could be interesting to incorporate into this model some ideas from the models presented previously.
For example, to incorporate the power-law distribution of attractiveness -- as done by Yakubo (see section~\ref{sec_yakubo}) -- or to use a range of interaction as captured by the parameter $\gamma$ discussed in the gravity models context (section~\ref{sec_gravity_simplest}).

\section{Summary \& Discussion}

The research of urban scaling attained a mature state where initial indication in data has been expanded to a solid empirical finding -- and to a range of theoretical models mathematically deriving the empirical exponents.
We review a set of modelling approaches, whereas most of the considered models (a) aim at explaining the emergence of non-linear urban scaling and (b) consist of a formal derivation.
The purpose of this review is, on the one hand, to summarize the models in a comprehensible and coherent manner and, on the other hand, to compare and relate them in order to identify similarities and dissimilarities.

We propose the taxonomy depicted in Fig.~(\ref{Fig_tax}).
The first distinguishing property is that 
the models can be divided into two types: the ones based on processes or interactions within a city and the ones based on processes between cities.
In the former, intra-city type, most models consider that urban scaling emerges from human interactions, i.e.\ intentionally or accidentally meeting people, in the considered city.
E.g.\ the model proposed by Bettencourt \cite{bettencourt2013} employs an analogy to the cross-section as known from physics. 
Other authors propose other mechanisms based on required human interactions within cities, and we identify a set of models employing gravity ideas.
To the other branch, the intra-city type, we count three papers that elaborate on very different ideas.
E.g.\ the work by Ribeiro et al.\ \cite{HRibeiro2021} argues in favour of a link between urban scaling and city size distribution.

We find that models employing human interaction within a city are based on a given probability of interaction.
This motivates us to unify this conceptual overlap in a framework that formalizes this probability. 
Indeed the models of this group only differ in the reasoning behind how the probability of interaction is estimated theoretically. The framework also includes models employing gravity ideas.

Reviewing the gravity models, we find that three groups of authors independently employ different ideas that lead to equivalent results (within given parameter configurations).
We interpret this observation as strong support for the gravity idea, i.e.\ that some sort of interaction decays as a power-law with some sort of distance, interfering in some way the urban scaling. 
A closer inspection of the variants permits us to draw valuable conclusions in terms of (i) good access to all parts of the city, (ii) influencers reaching distant parts of the city, and (iii) interaction between socially distant people. We also discuss how these aspects affect the scaling and socio-economic development of cities.
The analogy to gravity in physics has a long history and has been studied in a wider context, including population flows \cite{Simini2012}, like commuters \cite{Spadon2019}, or spatial explicit modelling \cite{Rybski2013}. 
Accordingly, as a small outlook, it would be interesting to further unify gravity applications which could then also lead to an explanation of the $\gamma$-exponent itself, that is, the parameter that controls the space impedance.

We also review models based on processes between cities and find that this school of thought is promising but much less developed.
First, there are fewer models in this branch.
Second, such models are less explicit, and the emergence of urban scaling is barely derived but, e.g.\ rather related to other scaling laws.
Still, we argue that cities are not isolated objects and that there are important processes between them.
At the same time also intra-city models are based on reasonable assumptions and prove to be successful in describing urban scaling.
It is reasonable to think that models that incorporate both
intra- and inter-city processes could represent an important step towards a \textit{Unified Urban Theory} (UUT) \cite{Ribeirofisica2020,Lobo2020,Bettencourt2010}

To finalize, it is important to mention other models and approaches published recently that also help understanding the urban scaling properties but are somewhat different from the models presented here.
For instance, some works that analyzed the urban scaling as a consequence of other power-law functions/distributions, as is the case of Bettencourt's energy dissipation model, discussed in the second part of his paper \cite{bettencourt2013}.
This kind of approach is important to describe urban scaling; however, it does not explain or derive the scaling exponent more fundamentally, as do the models here presented. 
Another group of works also dedicated to explain/determine the scaling exponent employs data-driven approaches, as is the case of \cite{Pan2013,Dong2020,louf2014congestion,Bettencourt2020a}.

We also need to mention that here we exclusively consider urban scaling cross-sectionally.
The equivalence of cross-sectional (transversal) and temporal scaling (longitudinal) is tempting but requires some formal considerations.
Recently, two groups independently and consistently provide the theoretical description \cite{Ribeiro2020, Bettencourt2020}.
Essentially, these works suggest that temporal and cross-sectional scaling are the same, given that cities have sufficiently large growth rates and without exogenous factors. 
Rest for future works analyze the connection between the models here presented in the context of individual cities growth.

Overall, we hope this paper disentangles the plethora of urban scaling models and thereby reveals similarities and dissimilarities.
Substantial progress has been made by the community, but we also think that the chapter of urban scaling models cannot be closed yet.

\begin{acknowledgments}
We wrote this paper in memoriam of Joao Meirelles, who unexpectedly passed away in 2020 and who helped with the first insights that gave rise to this work.
We would like to thank the various authors who supported us by verifying the description of their models and approaches.
F. L. Ribeiro thanks  CAPES (grant number 88881.119533/2016-01), CNPq (grant number 405921/2016-0), and Fapemig.
D.\ Rybski thanks the Alexander von Humboldt Foundation for financial support under the Feodor Lynen Fellowship.
D.\ Rybski is grateful to the Leibniz Association (project IMPETUS) for financially supporting this research. \end{acknowledgments}


\end{document}